\documentclass[letterpaper,12pt]{article}

\usepackage{tabularx} 
\usepackage{jheppub}
\usepackage{amsmath}  
\usepackage{graphicx} 
\usepackage{subcaption}
\usepackage[utf8]{inputenc}
\usepackage[T1]{fontenc}
\usepackage{lmodern}
\usepackage{tikz}

\usepackage{hyperref}
\def\be{\begin{equation}}
\def\ee{\end{equation}}

\begin{document}

\title{An Analytic Zeta Function Ramp \\ at the Black Hole Thouless Time}
\author[a]{Pallab Basu,}
\author[a]{Suman Das,}
\author[b]{Chethan Krishnan}
\affiliation[a]{Mandelstam Institute for Theoretical Physics, School of Physics, University of the Witwatersrand, Johannesburg, WITS 2050, South Africa.}
\affiliation[b]{Center for High Energy Physics,
Indian Institute of Science, Bangalore 560012, India}
\emailAdd{pallab.basu[at]wits.ac.za, suman.das[at]wits.ac.za, chethan.krishnan.physics[at]gmail.com}
\date{\today}


\abstract{Black hole normal modes have intriguing connections to logarithmic spectra, and the spectral form factor (SFF) of $E_n = \log n$ is the mod square of the Riemann zeta function (RZF). In this paper, we first provide an analytic understanding of the dip-ramp-plateau structure of RZF and show that the ramp at $\beta \equiv \Re(s)=0$ has a slope precisely equal to 1. The $s=1$ pole of RZF can be viewed as due to a Hagedorn transition in this setting, and Riemann's analytic continuation to $\Re(s)< 1$ provides the quantum contribution to the truncated $\log n$ partition function. This perspective yields a precise definition of RZF as the ``full ramp after removal of the dip'', and allows an unambiguous determination of the Thouless time. For black hole microstates, the Thouless time is expected to be $\mathcal{O}(1)$--remarkably, the RZF also exhibits this behavior. To our knowledge, this is the first black hole-inspired toy model that has a demonstrably $\mathcal{O}(1)$ Thouless time. In contrast, it is $\mathcal{O}(\log N)$ in the SYK model and expected to be $\mathcal{O}(N^{\#})$ in supergravity fuzzballs. We trace the origins of the ramp to a certain reflection property of the functional equation satisfied by RZF, and suggest that it is a general feature of $L$-functions--we find evidence for ramps in large classes of $L$-functions. As an aside, we also provide an analytic determination of the slopes of (non-linear) ramps that arise in power law spectra using Poisson resummation techniques.}

\maketitle

\section{Introduction}

Black holes are peculiar as quantum mechanical systems in that it was their quasi-normal modes that were computed first \cite{Vishveshwara:1970zz}, and not their normal modes \cite{tHooft, FirstRamp}\footnote{The results of \cite{tHooft, FirstRamp} can be viewed as increasingly explicit, but crude,  determinations of black hole normal modes. See also various follow ups in \cite{FuzzRandom, SimplestRamp, SumanArnab, Pradipta1, Burman1,Souvik, Ranjini, Suchetan, Burman2, SumanBaishali, Souvik2, Pradipta2}.}. This is both extremely natural and deeply problematic. It is natural because we expect the horizon to be a smooth place in bulk effective field theory at least in the large-$N$ limit, and therefore the natural boundary conditions are infalling. But it is also problematic, because it cannot be the case that quasinormal modes are the true modes of a unitary quantum mechanical system. They have imaginary parts and they result in eternal (in particular post-Page) decay and information loss.

In order to understand the microstates of generic black holes, it seems necessary to have control over quantum gravity beyond our present abilities. In full theories of quantum gravity like $\mathcal{N}=4$ SYM, it is simply not clear how one might compute the normal modes of heavy states that are dual to black holes. The best one can do typically is to consider toy models like the SYK model to mimic aspects of these heavy, highly chaotic states. 

On general grounds, it is believed that the spectrum of black hole microstates should exhibit aspects of random matrix (RMT) behavior \cite{Cotler, ShenkerBPS}. We will emphasize four RMT-like features that are expected of black holes:
\begin{itemize}
\item The spectral form factor (SFF) has a linear ramp of slope $\sim 1$ on a log-log plot.
\item The ramp has fluctuations and is not smooth due to averaging.
\item Spectrum has level repulsion.
\item The Thouless time is $\mathcal{O}(1)$. 
\end{itemize}
Some of these terms will be elaborated on in later sections, see \cite{Cotler, ShenkerBPS} for more details.

In \cite{Pradipta1, Burman1, Burman2}, it was noted that if one installs a brick wall or a stretched horizon at a Planckian distance from the horizon as a UV regulator, the excitations of the quantum fields with these boundary conditions are essentially indistinguishable for an exterior observer, from those of a smooth horizon. This includes the correct thermodynamics, as well as exterior correlators. The latter are exponentially close (in the black hole entropy) to smooth horizon correlators. In papers to appear in the near future, we will strengthen these conclusions further\footnote{In particular, one of the unsatisfying features of the calculations in \cite{Burman1,Burman2} was that one needed to introduce a cut-off in the angular quantum number $J$, by hand. We have now understood that this has a physical explanation related to the angular momentum effective potential and the weak $J$-dependence of the spectrum.}. 

If one's goal is only to reproduce the correct thermodynamics and time-order correlators, it turns out that one can ignore the (weak) $J$-dependence of the black hole normal modes \cite{Burman1,Burman2}. But instead if one decides to retain the $J$-dependence, these normal modes can be used to go beyond mere thermal properties and capture aspects of the chaotic dynamics. In particular, they were shown to reproduce the first three of the four bullet points listed above, in \cite{FirstRamp, FuzzRandom, SimplestRamp}. 

Despite the fact that they give rise to RMT like features, in the low-lying part of the spectrum which is responsible for black hole physics, these normal modes have an approximately logarithmic dependence on the angular quantum numbers (which we have schematically denoted here by $J$). As we will see in great detail, the partition functions and SFFs of the logarithmic spectrum are closely related to the Riemann zeta function (RZF). It was shown numerically in \cite{SimplestRamp} that the linear ramp is in fact reflected in the RZF. The fact that deterministic sequences like $\log n$ (and not just random matrices) can give rise to linear ramps lead \cite{SimplestRamp} to consider other deterministic sequences, interpret them as eigenvalues of a Hamiltonian, and investigate their spectral features. Some of them exhibited non-linear ramps. 

\subsection{Deterministic Sequences, Chaos and the Thouless Time}

In \cite{FirstRamp}, it was first shown that the spectral form factor (SFF) constructed from the single-particle partition function of the normal modes of the BTZ black hole (with a brick wall) exhibits a dip-ramp-plateau structure, with a linear ramp of unit slope. This provided the first example of a spectrum drastically different from that of a random matrix, with a linear ramp in the SFF. Notably, these modes can be thought of approximately as a deterministic sequence, which is fundamentally different from the spectrum of a chaotic system. This observation raises a general question: when does a ramp appear in a deterministic sequence when interpreted as the spectrum of a Hamiltonian? This question was partially addressed in \cite{SimplestRamp}, where it was shown that a dip-ramp-plateau structure with a ramp of constant slope appears in many deterministic sequences (truncated at some point). A key criterion for the existence of the ramp was identified as  
\begin{equation}
    \lim_{n\rightarrow \infty} \log_{n} E_n <1.
\end{equation}  
Table 1 of \cite{SimplestRamp} lists several sequences that exhibit a ramp in the SFF. Among these, a particularly simple class of sequences is given by \( E_n = \{n^\alpha\} \). It was observed that for \( \alpha > 1 \), no clear ramp appears, whereas in the regime \( \alpha < 1 \), a ramp of constant slope emerges. As \( \alpha \) decreases, the slope also decreases, approaching unity in the limit \( E_n = \log n \), which can be interpreted as the \( \alpha \to 0 \) limit of the sequence. Moreover, if \( \alpha \) is further decreased beyond zero, the slope continues to decrease and eventually vanishes at some point (see Figure 1 of \cite{SimplestRamp}). Numerical fitting suggests that the slope of the ramp for the sequence \( E_n = n^{\alpha} \) is approximately given by  $1/(1-\alpha)$.

Perhaps most interesting from both the physics and the mathematics perspectives is the log-spectrum $E_n = \log n$. From physics, the interest in it lies in what we mentioned earlier -- that the log spectrum has close connections to black hole normal modes. From mathematics, what makes the log spectrum interesting is that the (complexified) partition function in this case is  the Riemann zeta function $\zeta(s)$. It is easy to check that the RZF exhibits a ramp in the SFF. Strong numerical evidence that this ramp has a slope of 1 was presented in \cite{SimplestRamp}.

Our orienting goal in this paper is to develop an analytic understanding of the ramps in various sequences, with a particular focus on the logarithmic spectrum. We will show analytically, using properties of the zeta function, that the slope of the SFF of the \( \log n \) spectrum is indeed one. Related observations lead us to a rich picture of the RZF in terms of a Hagedorn transition and a natural interpretation of the dip and the ramp as classical and quantum contributions to the SFF of the log spectrum. We are also able to make statements about the plateau in terms of a ``dynamical'' phase transition (to be distinguished from a static or ordinary phase transition which is a statement about the partition function without an imaginary part in the inverse temperature). We also observe that many of these observations naturally generalize to so-called $L$-functions. These are number theoretic functions that are believed to be generalizations of RZF -- we will discuss them in some detail in Section \ref{sec:dirichlet_sff} and some of the appendices. Towards the end of the paper is Section \ref{sec:OtherSeq}, we will also demonstrate, using a Poisson resummation formula, that the slope of the \( E_n = \{n^{\alpha}\} \) spectrum is given by \( 1/(1-\alpha) \), as guessed numerically in \cite{SimplestRamp}.  

One of the lessons that emerges from our work is that it may be natural to view the pole at $s=1$ of the RZF as due to a Hagedorn transition (in the primon gas, see Appendix). The RZF can be viewed as the quantum correction to the partition function of the log spectrum truncated at integer $N$. This $N$ is the size of the system and $1/N$ is a measure of the quantum corrections. The classical contribution to the SFF is essentially the dip, and the SFF of the RZF gets a precise definition as the ``ramp-after-removal-of-the-dip''. The beginning time of the ramp after removal of the dip is often used as a definition of the Thouless time\footnote{The latter is viewed as a measure of the strength of chaos -- the smaller the Thouless time in a system, the stronger the mixing. Black holes, being strong mixers, are expected to have an $\mathcal{O}(1)$ Thouless time.}. This means that we can determine the Thouless time of the log $n$ spectrum for any $N$, from the RZF ramp. We find that the Thouless time is $\mathcal{O}(1)$ for the RZF, a feature expected for black holes \cite{ShenkerBPS} as well. In fact, as far as we are aware, this is the first model with connections to black holes that has an $\mathcal{O}(1)$ Thouless time, apart from random matrices. We will discuss these observations in more detail in a later section.

\section{Spectral Form Factor and the Dip-Ramp-Plateau}\label{SFF_DRP}

The spectral form factor (SFF) is defined as
\begin{equation}
g(t, \beta) = Z(\beta + i t) Z(\beta - i t).
\end{equation}
Sometimes, $g(t,\beta)$ is normalized by $Z(\beta)^2$ \cite{Cotler}. Here, $Z(\beta+it)$ is the analytically continued partition function of the system of interest at inverse temperature $\beta \rightarrow \beta + i t$.

At very early times, the discreteness of the spectrum is not important and can be coarse-grained by approximating the spectrum as a continuous function and the sum in the partition function as an integral. We refer to this as the classical contribution. Thus, at any time, $Z(s=\beta+it)$ can be written as an integral that gives the classical contribution plus some additional terms, which we call quantum corrections, as follows:
\begin{align}\label{classquant}
    Z(s;{E_n}) &= \sum_{n=1}^{\infty} \exp[-s E_n] \nonumber \\
    &= \int \exp[-s E(x)] dx + \text{quantum corrections} \nonumber \\ 
    &= Z_{cl}(s;E(x)) + \text{quantum corrections}.
\end{align}
At late times, the classical part typically decays due to the gapless nature of the classical system, allowing the quantum contribution to dominate in the SFF. We refer to this decaying part as the \textit{dip}. At very late times, the discreteness of frequencies leads to a characteristic \textit{plateau} behavior ($g(t,\beta) \sim \text{finite constant}$). To illustrate this, let us express the SFF (focusing on $\beta=0$ for simplicity) as:
\begin{equation}
g(t, \beta=0) = \sum_{m,n=1}^{N} e^{i(E_m - E_n)t}.
\end{equation}
At late times, the phases become highly oscillatory and average out to zero, except for the contributions where $E_m=E_n$. The long-time average is therefore:
\begin{equation}\label{plateau0}
\lim_{T \to \infty}\frac{1}{T} \int_{0}^{T} g(t, \beta=0) dt = \sum_{m=1}^{N} 1 = N.
\end{equation}
This implies that $g(t, \beta=0)$ saturates to a plateau whose height scales as $N$, with oscillations superimposed on top. Related arguments apply for nonzero $\beta$ as well \cite{Cotler}, we will discuss them later for the cases of interest.

These two features—the initial dip and the plateau—are generic to quantum systems with discrete spectra. The nontrivial aspect is how the dip connects to the plateau. In some systems, these two regimes are linked by a \textit{ramp}-like behavior, where $g(t,\beta) \sim t^{\alpha}$. For chaotic systems, it is observed that $\alpha = 1$. Furthermore, in random matrix theory (RMT), it can be shown analytically that this exponent is precisely one \cite{Mehta1960ONTS, dyson1962statistical, Cotler}.

\begin{figure}
    \centering
    \includegraphics[width=.55\textwidth]{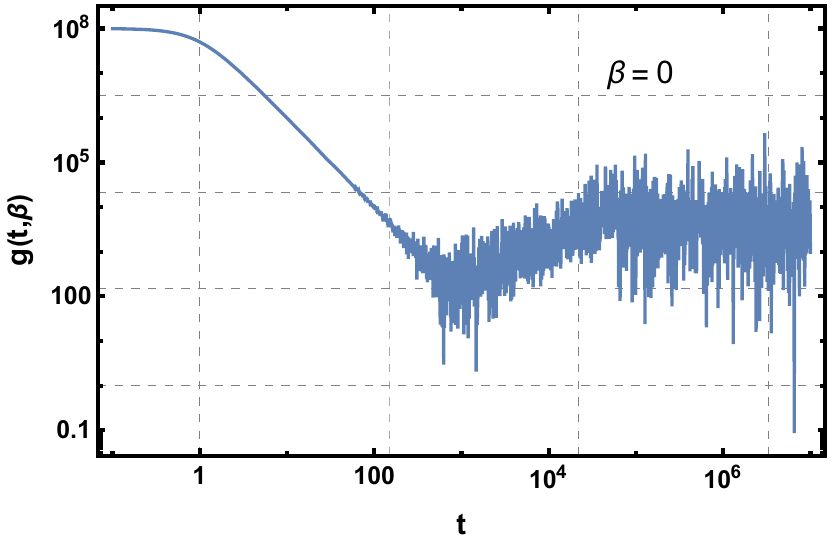}
   \caption{Log-log plot of the SFF $Z_N(\beta=0)$ constructed from the truncated log spectrum with $N=10,000$. This SFF shows a clear dip, ramp, and plateau at late times. We will later show analytically that the slope of the ramp is precisely 1.}
    \label{SFF_logN}
\end{figure}

We will be interested in deterministic spectra (and not directly RMT) in this paper. Figure \ref{SFF_logN} shows the SFF behavior for energy levels given by $E_n = \log n$, truncated at $n_{max} \equiv N = 10000$. The primary goal of this figure is to show that the dip-ramp-plateau structure is {\em not} necessarily a feature only of conventional RMT (despite the lore to the contrary).

%
%

\section{The Physics of the Log Spectrum}

Let us first focus on the logarithmic spectrum $E_n=\log n$, which has been studied previously due to its connection to the primon gas \cite{JuliaNumbers, BakasGases}. The partition function constructed at inverse temperature $\beta$ is
\begin{align}\label{partition0}
    Z_N(\beta) &= \sum_{n=1}^{N} e^{-\beta \log n} = \sum_{n=1}^{N} \frac{1}{n^{\beta}}
\end{align}
For $\beta>1$ the sum is convergent and we can do the full sum which gives zeta function, i.e,
\begin{equation}
    \lim_{n\rightarrow \infty} Z_N(\beta)=Z_{\infty}(\beta)=\zeta(\beta).
\end{equation}
This is, in fact, the grand canonical partition function of a non-interacting bosonic primon gas at inverse temperature $\beta$ (see appendix \ref{AppPrimon} for details). However, the partition function diverges for $\beta \leq 1$, indicating that $\beta = 1$ plays a special role. In the subsections that follow, we will explore this point in more detail, along with the behavior of the partition function and SFF in the regime $\beta \leq 1$.

\subsection{Hagedorn Transition: A Static Phase Transition}

Since our primary interest is the SFF, which involves complexified $\beta$, let's denote $s=\beta+it$ so that $\Re(s)=\beta$. The partition function does not converge for $\Re(s)<1$. To examine the divergence for $\Re(s)<1$, we approximate the series sum as an integral:
\begin{align}
    Z_N(s)=\int_{1}^N \frac{dx}{x^s} &= \frac{N^{1-s}}{1-s}-\frac{1}{1-s}=
    \begin{cases}
  \mathcal{O}(1), &  \beta>1 \\
  \text{diverges}, & \beta\leq1.
\end{cases}
\end{align}
Thus, for $\beta>1$, $Z_N(s)$ remains finite and $O(1)$, whereas at $s=1$, it diverges and becomes large for $\beta<1$ as $N\rightarrow \infty$. Consequently, the untruncated partition function $Z_{\infty}(s)=\zeta(s)$ develops a simple pole at $s \rightarrow1^{+}$ \footnote{The existence of a simple pole of $\zeta(s)$ can be infer by expanding $\zeta(s)$ as Laurent series near $s =1$. To obtain this, we use the Euler-Maclaurin formula, which approximates sums by integrals,
\begin{equation}
\sum_{n=a}^{b} f(n) \approx \int_a^b f(x) \,dx + \frac{f(a) + f(b)}{2} + \sum_{k=1}^{\infty} \frac{B_{2k}}{(2k)!} f^{(2k-1)}(a),
\end{equation}
where $B_k$ are the Bernoulli numbers. Applying this formula to $\zeta(s)$, we obtain: \begin{equation}
\zeta(s) = \frac{1}{s-1} + \sum_{k=0}^{\infty} \frac{(-1)^k}{k!} \gamma_k (s-1)^k,
\end{equation}
where $\gamma_k$ are the Stieltjes constants. This shows that $\zeta(s)$ has a simple pole at $s=1$.},
\begin{equation}
    Z_{\infty}(s) \sim \frac{1}{s-1}.
\end{equation}
This behavior is reminiscent of large-$N$ gauge theories, particularly $\text{SU}(N)$ gauge theories \cite{witten1998antidesitterspacethermal}, where the free energy scales as  
\begin{align}
    F \sim  \begin{cases}
  \mathcal{O}(1), & \quad \text{for } \beta > \beta_H \quad \text{(confined phase)} \\
  \mathcal{O}(N^2), & \quad \text{for } \beta < \beta_H \quad \text{(deconfined phase)}.
\end{cases}
\end{align}
Similar to the zeta function case, the infinite-cutoff ($N \to \infty$) partition function exhibits a pole at $\beta = \beta_H$.  

The physical origin of the Hagedorn transition \cite{Aharony:2003sx,Sundborg:1999ue} in large-$N$ theories is worth reviewing. While standard state counting suggests a density of states  $\rho(E) \sim \exp(\beta_H E)$, the large-$N$ framework introduces crucial modifications when energies scale with $N$ \cite{witten1998antidesitterspacethermal}. At the threshold $E \sim N^2$, non-planar diagrams and multi-trace operators become dominant, effectively acting as a cutoff on the exponential Hagedorn growth. This suppression mechanism is thermodynamically significant—rather than leading to an infinite energy density above the Hagedorn temperature, it triggers a phase transition to a deconfined phase characterized by a fundamental reorganization of degrees of freedom.  

The deconfined phase is often associated with a black hole in the dual theory. Consequently, we expect the spectral form factor (SFF) to exhibit chaotic behavior \cite{Cotler}.  We will usually be interested in the post-Hagedorn (deconfined) phase in what follows, because below the Hagedorn temperature the partition function of the log spectrum converges to the RZF. On the other hand, we will see that in the deconfined phase, the RZF has an interpretation as the quantum correction to the classical dip.

\subsection{The Dip as a Classical Contribution}

\begin{figure}
    \centering
    \includegraphics[width=.55\textwidth]{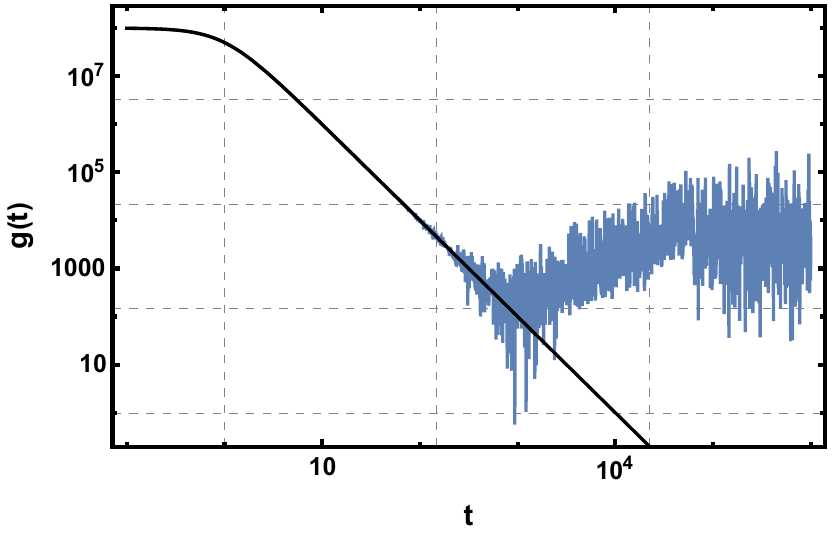}
   \caption{SFF as in Figure \ref{SFF_logN}, but with equation \eqref{dip2} plotted in black on top, showing that the dip part comes from the first term of Equation \eqref{truncated_zeta}.}
    \label{Dip_fit}
\end{figure}

In Figure \ref{SFF_logN}, we observed the behavior of the spectral form factor (SFF) over time, constructed from $Z_N(\beta=0)$, revealing a dip-ramp-plateau structure similar to that observed in chaotic systems. To understand how this structure arises in $Z_N$, let us consider the following relation \cite{Titchmarsh}:
\begin{equation}
\zeta(s) = \sum_{n=1}^{N} \frac{1}{n^s} - \frac{N^{1-s}}{1 - s} + \frac{1}{2} N^{-s} + \sum_{k \geq 1} \frac{B_{2k}}{(2k)!} \frac{\Gamma(s+2k-1)}{\Gamma(s)} N^{-s-2k+1} ,
\end{equation}
where $B_{2k}$ are the Bernoulli numbers. For large $N$, we can retain only the first term in the last summation. Rewriting this, we obtain:
\begin{equation} \label{truncated_zeta}
 Z_N(s) \approx \frac{N^{1-s}}{1 -s}+ \zeta(s) + \frac{1}{2} N^{-s} - \frac{s}{12}  N^{-1-s}.
\end{equation}
This expression closely resembles equation \eqref{classquant}, where the first term represents a coarse-grained classical approximation, while the remaining terms correspond to quantum corrections. To make this interpretation clearer, we approximate the partition function in terms of the density of states $\rho(E)$ as follows:
\begin{equation}
Z = \sum_{n=1}^{N} e^{-\beta E_n} \approx \int \rho(E) e^{-\beta E} dE.
\end{equation}
In \eqref{classquant}, this integral was identified as the classical contribution $Z_{cl}$. For the log spectrum, using $\rho(E) = dn/dE$, this can be written as:
\begin{align}
Z_{cl}(\beta) = \int_{1}^{N} \frac{dn}{n^{\beta}} \
= \frac{N^{1-\beta} - 1}{1 - \beta} \approx \frac{N^{1-\beta}}{1 - \beta}.
\end{align}
where in the last step, we have taken $\beta <1$.
Thus,
\begin{align}\label{dip_part}
Z_{cl}(\beta+i t)Z_{cl}(\beta-it) = \frac{N^{2-2 \beta }}{(1-\beta)^2+t^2}.
\end{align}
Consequently, the classical part of the SFF (commonly referred to as the disconnected part) becomes:
\begin{equation}\label{dip2}
g_{cl}(t) = \frac{N^{2(1-\beta)}}{(1-\beta)^2+t^2} = \frac{N^{2(1-\beta)}}{t^2} + O(t^{-4}).
\end{equation}
Clearly, this differs from the well-known $t^{-3}$ decay of the dip in random matrix theory (RMT) SFF. In Figure \ref{Dip_fit}, we have plotted the SFF for $\beta=0$ along with $N^2/(1+t^2)$ in black. This illustrates that the dip in the SFF originates from the classical approximation, corresponding to the first term on the right-hand side of Eq.~\eqref{truncated_zeta}.
%
%

\subsection{The Ramp and Plateau as Quantum Corrections}

\begin{figure}
    \centering
    \includegraphics[width=.55\textwidth]{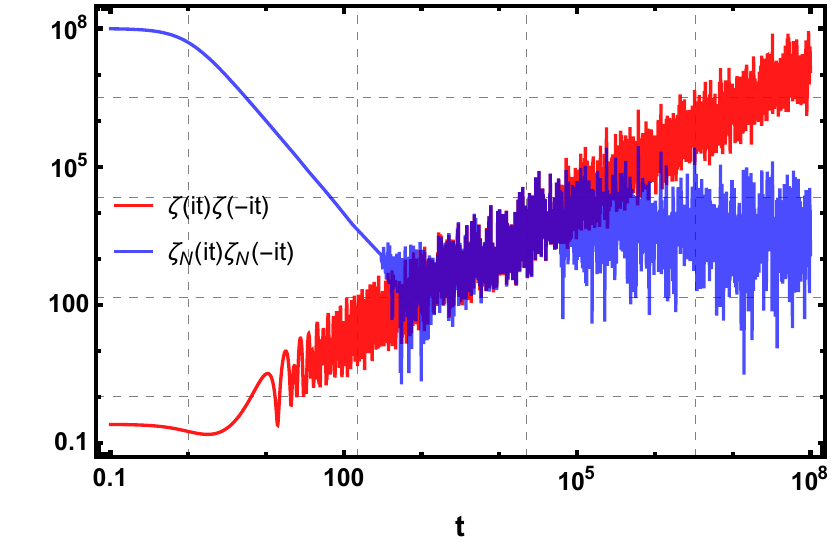}
   \caption{In this figure, we present the SFF of the logarithmic spectrum truncated at \( N = 10,000 \) (blue) alongside \( \zeta(it) \zeta(-it) \) (red). This illustrates that the ramp in the SFF arises from the second term of \eqref{truncated_zeta}. }
    \label{ramp_fit}
\end{figure}

Once the effect of the first term decays, the quantum corrections come into play and most important of them is the second term $\zeta(s)$ because $\zeta(it)\zeta(-it)$ exhibits an ever-increasing ramp as shown in red in Figure \ref{ramp_fit}. This Figure clearly shows that the ramp in the SFF comes from the second term of \eqref{truncated_zeta}.

Determining which term in Eq. \eqref{truncated_zeta} gives rise to the plateau is challenging because the plateau occurs at very late times, where our approximations break down. However, we can infer the presence of the plateau using the argument presented in Eq. \eqref{plateau0}. There are, however, some subtleties, which we discuss below in detail.

\subsection{Plateau Height and a Dynamical Phase Transition}

The SFF at non-zero $\beta$  can be expressed in terms of the eigenvalues of the Hamiltonian as  
\begin{align}\label{sff_00}
    g(t, \beta) &= \sum_{m, n} e^{-(\beta+it)E_m} e^{-(\beta-it)E_n} \nonumber \\
    &= \sum_{m, n} e^{-\beta(E_m+E_n)} e^{-it(E_m-E_n)}.
\end{align}
As discussed in Section \ref{SFF_DRP}, at very late times, the phase factor in \eqref{sff_00} becomes highly oscillatory, averaging out to zero, with the only nonzero contributions coming from terms where \( E_m = E_n \). Consequently, the late-time average of the SFF is  
\begin{align}
     \langle g(t, \beta) \rangle &= \lim _{T\rightarrow \infty} \int _0^T g(t, \beta) dt \nonumber \\ 
     &\sim \sum _{n} e^{-2 \beta E_n} = \sum_{n=1}^N \frac{1}{n^{2 \beta}} \nonumber \\  
     &\approx \int_{1}^N \frac{dx}{x^{2\beta}} = \frac{N^{1-2\beta}}{1-2\beta} + O(1).
\end{align}
This result implies that the average plateau height is given by  
\begin{align}
    h_{\text{plateau}} = \frac{N^{1-2\beta}}{1-2\beta} + O(1),
\end{align}
but we need to be careful about the value of $\beta$.
The behavior of \( h_{\text{plateau}} \) depends crucially on the value of \( \beta \) and in the limit of the full untruncated log spectrum, we obtain  
\begin{align}
    h_{\text{plateau}} &= \zeta(2\beta), \quad \text{for} \quad \beta > \frac{1}{2}, \nonumber \\
    h_{\text{plateau}} &\rightarrow \infty, \quad \text{(no plateau)} \quad \text{for} \quad \beta < \frac{1}{2}.
\end{align}
This indicates a kind of phase transition in the SFF of the log-spectrum at \( \beta = \frac{1}{2} \). Unlike the Hagedorn transition, which is a \textit{static} phase transition (visible in the partition function), we refer to this as a {\em dynamical} phase transition (visible in the SFF).

\subsection{Analytic Origin of the Ramp}

As we saw in the previous section, $\beta=1$ and $\beta=1/2$ are special points, and for $\beta<1/2$, there is no plateau in the sense that the plateau height diverges. In this section, we provide alternate analytical arguments based on the properties of the RZF to explain these behaviors, as well as the linear ramp in the SFF.

Since the SFF exhibits fluctuations at late times, it is customary to perform some form of averaging, such as ensemble averaging in random matrix theory. However, for the logarithmic spectrum, the ensemble interpretation is not entirely clear. Instead, we consider time averaging to smooth out the fluctuations\footnote{Note that this is physical: ensemble-averaging in statistical mechanics often has an ergodic origin.}. We previously observed that the partition function of the logarithmic spectrum is related to the Riemann zeta function. Consequently, the time-averaged SFF can be expressed in terms of the mean value of the modulus squared of the zeta function:
\begin{equation}
     \langle g(T, \beta) \rangle=\lim_{T\rightarrow \infty} \frac{1}{T} \int  |\zeta(\beta+it)|^2 dt
\end{equation}
This expression will be our main focus and our goal is to analyze this expression using known mean value theorems of zeta function (see eg., \cite{Titchmarsh}) for different regimes of $\beta$.

\subsubsection{Case I: \(\beta > 1\)}

For $\beta>1$, the mean value theorem states that
\begin{equation}\label{betalarge}
    \lim_{T\rightarrow \infty} \frac{1}{T} \int _0^T |\zeta(\beta+it)|^2 dt =\zeta(2\beta)
\end{equation}
This implies that the time-averaged SFF saturates at a constant value, meaning a plateau exists with height given by Eq. \eqref{betalarge}.
\begin{figure}
     \centering
     \begin{subfigure}[b]{0.45\textwidth}
         \centering
         \includegraphics[width=\textwidth]{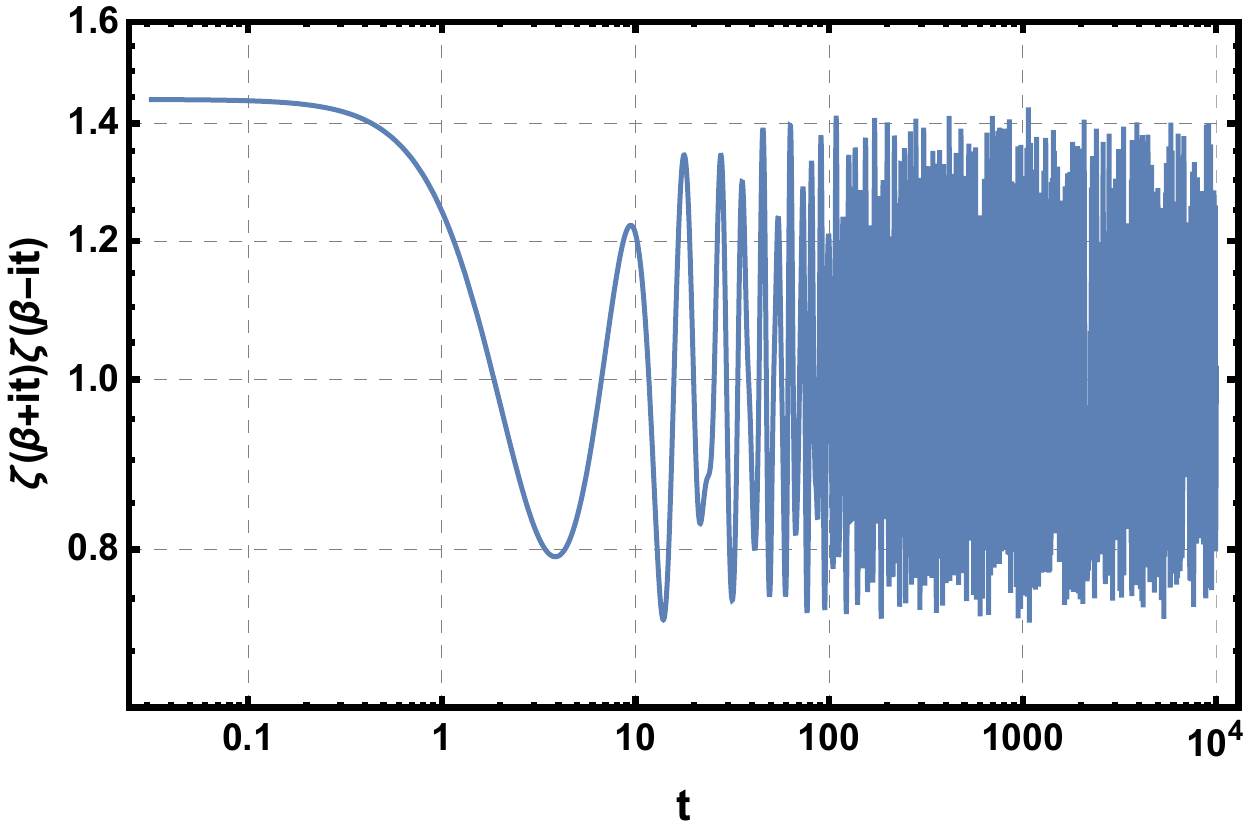}
         \caption{$\beta=3.0$}
         \label{fig_beta3}
     \end{subfigure}
     \hfill
      \begin{subfigure}[b]{0.45\textwidth}
         \centering
         \includegraphics[width=\textwidth]{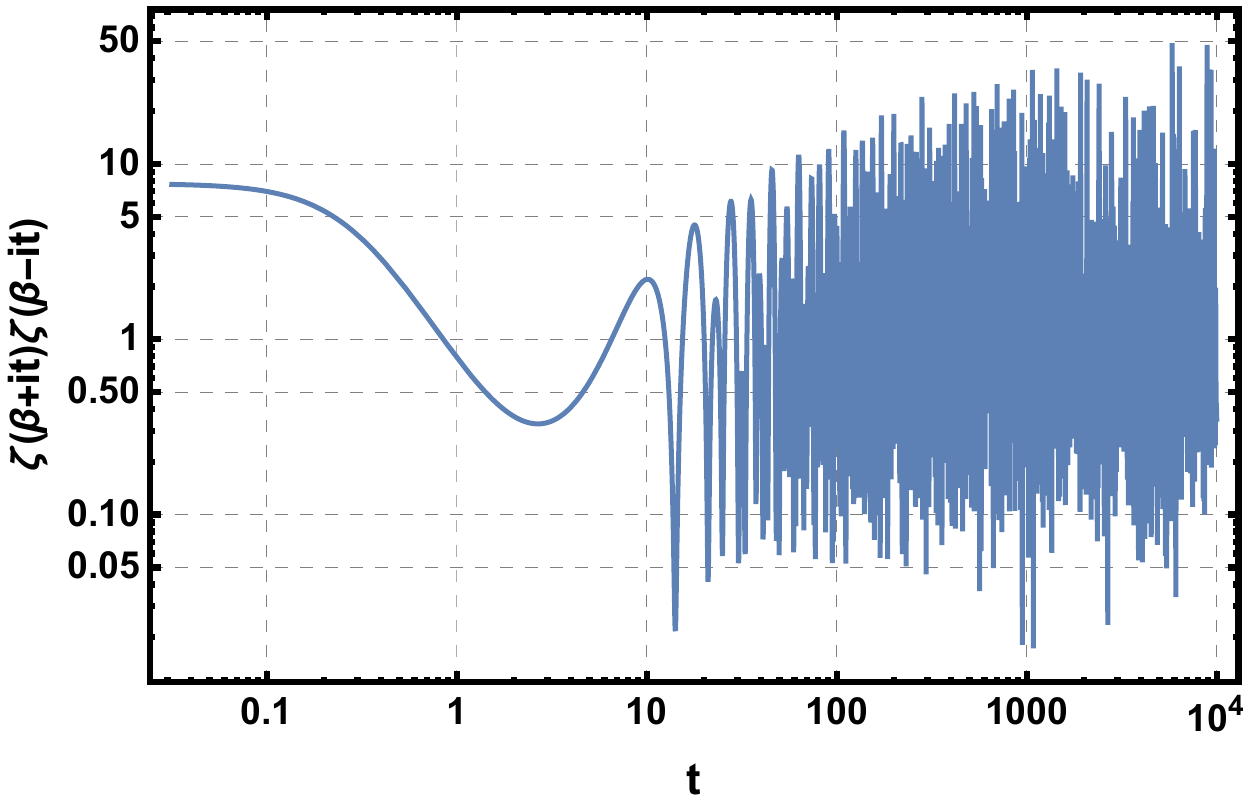}
         \caption{$\beta=0.7$}
         \label{fig_beta_1point7}
     \end{subfigure}
     \hfill
     \begin{subfigure}[b]{0.45\textwidth}
         \centering
         \includegraphics[width=\textwidth]{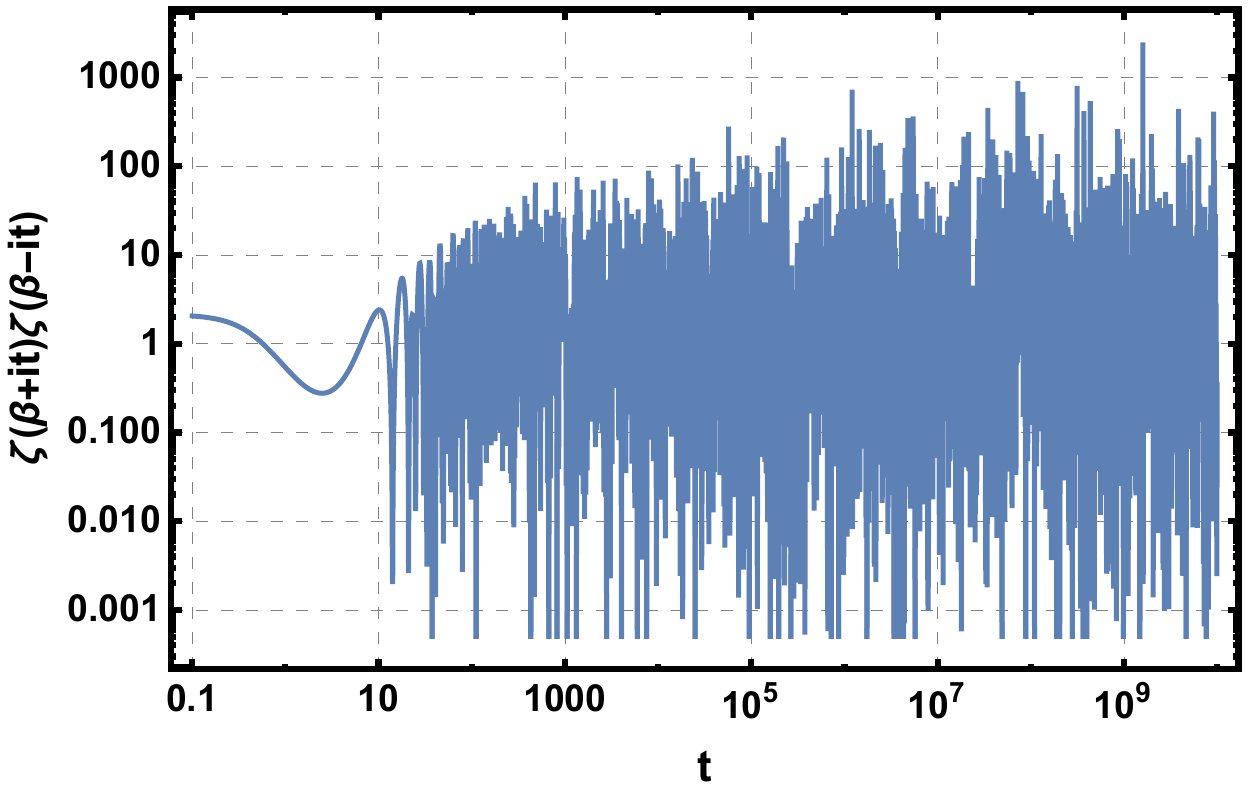}
         \caption{$\beta=0.5$}
         \label{fig_beta_half}
     \end{subfigure}
     \hfill
  \begin{subfigure}[b]{0.45\textwidth}
         \centering
         \includegraphics[width=\textwidth]{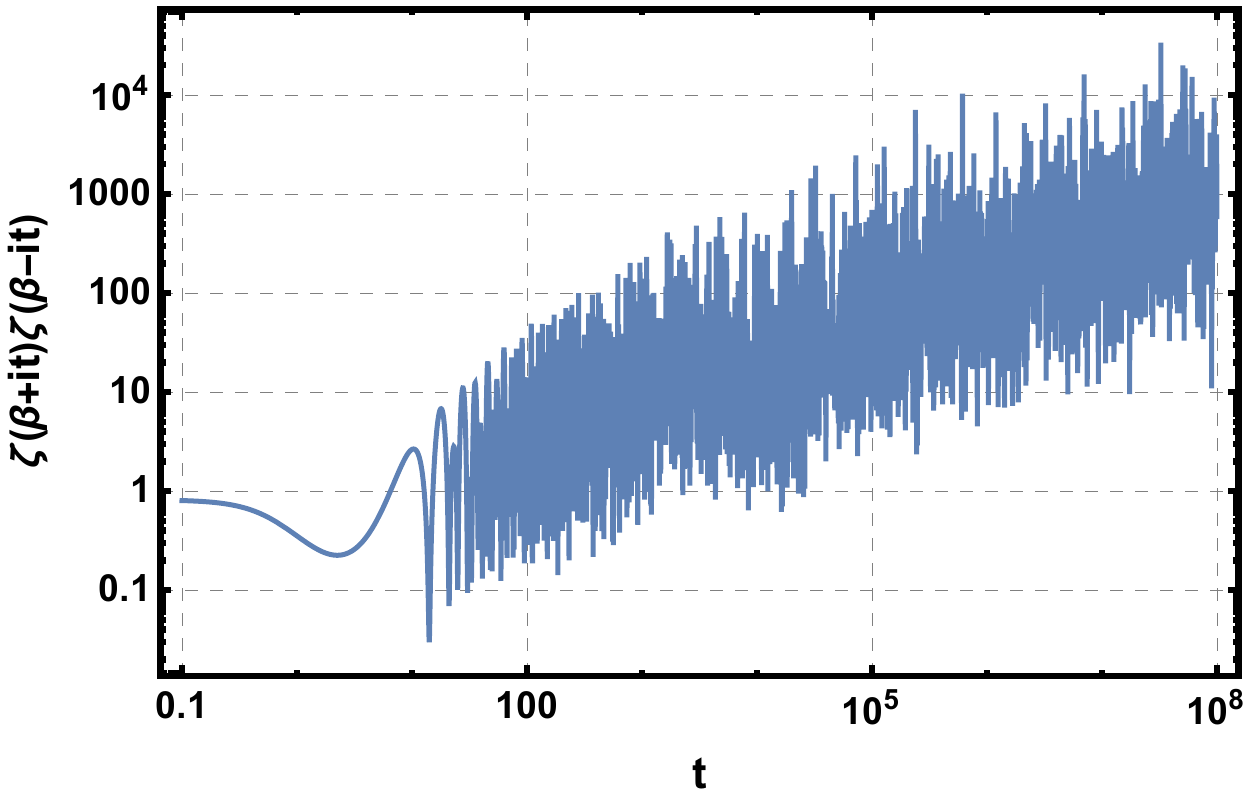}
         \caption{$\beta=0.3$}
         \label{fig_beta_point3}
     \end{subfigure}
        \caption{This set of figures illustrates the behavior of \( |\zeta(\beta + i t)|^2 \) for various values of \( \beta \). For \( \beta > \frac{1}{2} \), the function saturates (up to oscillations) at late times, showing no clear ramp. In contrast, for \( \beta < \frac{1}{2} \), it displays a persistent ramp with a well-defined slope proportional to \( t^{1 - 2\beta} \). At the critical point \( \beta = \frac{1}{2} \), the growth is logarithmic, though this is not prominently visible here (see Figure \ref{logfit_betahalf} for a clearer illustration). }
        \label{n_alpha_sff}
\end{figure}
%
%

\subsubsection{Case II: $\frac{1}{2}<\beta \leq 1$}

For 
$\frac{1}{2}<\beta \leq 1$, the mean value theorem gives
\begin{equation}\label{mean2}
    \lim_{T\rightarrow \infty} \frac{1}{T} \int _1^T |\zeta(\beta+it)|^2 dt =\zeta(2\beta)
\end{equation}
Thus, the time-averaged SFF again saturates at a constant value, implying the existence of a plateau with height given by Eq. \eqref{mean2}.

\subsubsection{Case III: $\beta =\frac{1}{2}$}

\begin{figure}
    \centering
    \includegraphics[width=.55\textwidth]{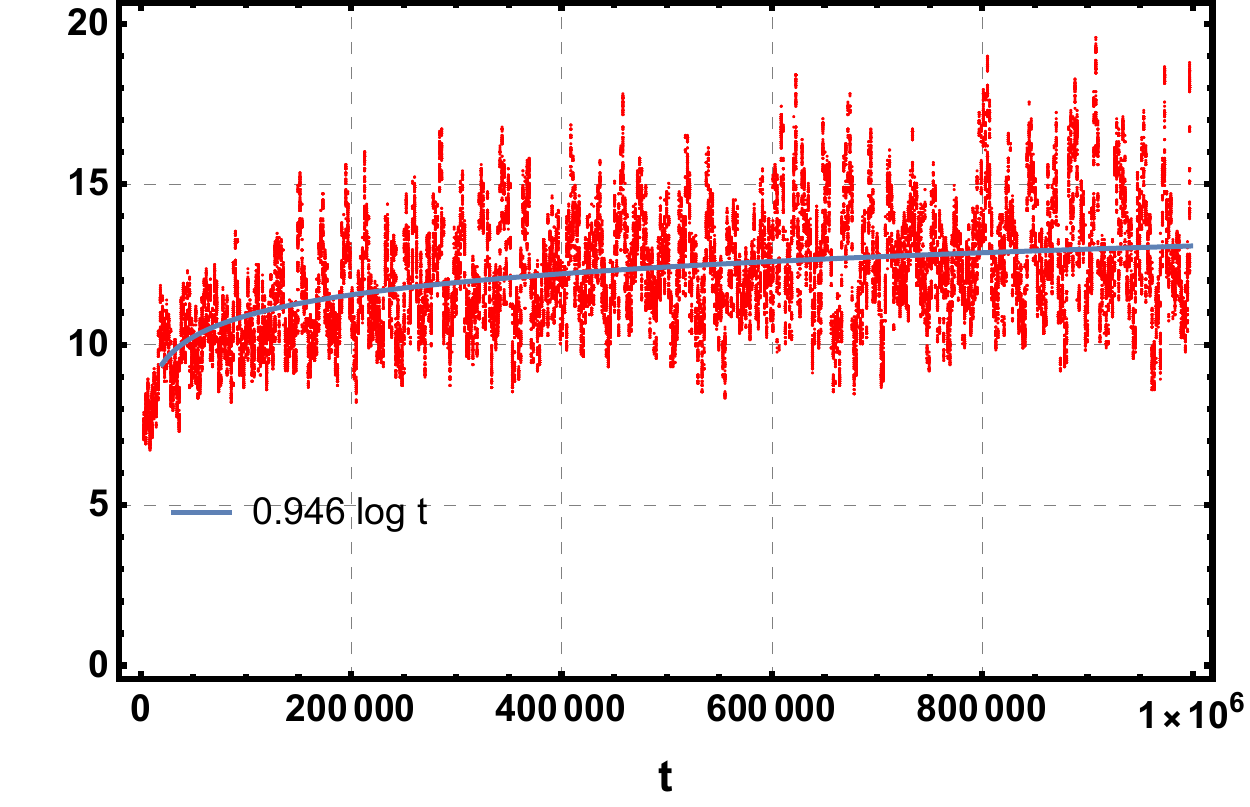}
   \caption{A moving-averaged version of $|\zeta\left(\tfrac{1}{2} + i t\right)|^2$ as a function of $t$ is plotted. The data is fitted with a logarithmic function, which is also shown in the figure.}
    \label{logfit_betahalf}
\end{figure}

For $\beta =\frac{1}{2}$, the mean value theorem states
\begin{equation}\label{loggrowth}
    \lim_{T\rightarrow \infty} \frac{1}{T} \int _1^T |\zeta(\beta+it)|^2 dt =\log T
\end{equation}
This result implies that at late times,
\begin{equation}\label{sff_betahalf}
    \langle g(t, \beta) \rangle \sim \log t, \quad \text{for} \ \  \beta= \frac{1}{2}.
\end{equation}
Since the SFF does not converge to a constant value, a well-defined plateau does not exist in this case. In Figure \ref{logfit_betahalf}, we have plotted the \texttt{MovingAverage} version of the SFF, i.e., the left-hand side of Eq.~\eqref{sff_betahalf}, with a window size of 400, shown in red. The blue curve represents the fitted function, which agrees with the logarithmic growth of the SFF.
%
%

\subsubsection{Case IV: \(0 \leq \beta < \frac{1}{2}\)}

\begin{figure}
\begin{subfigure}{0.51\textwidth}
    \centering
    \includegraphics[width=\textwidth]{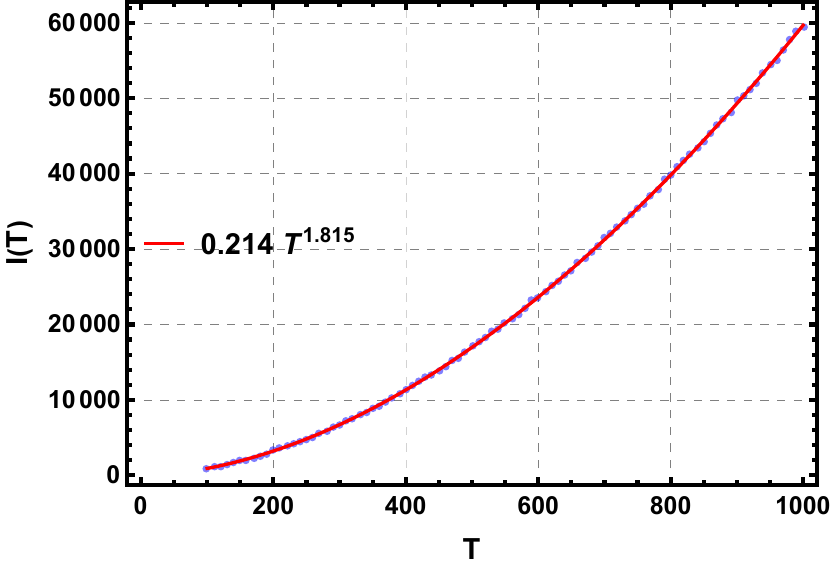}
    \end{subfigure}
    \hfill
    \begin{subfigure}{0.53\textwidth}
    \includegraphics[width=\textwidth]{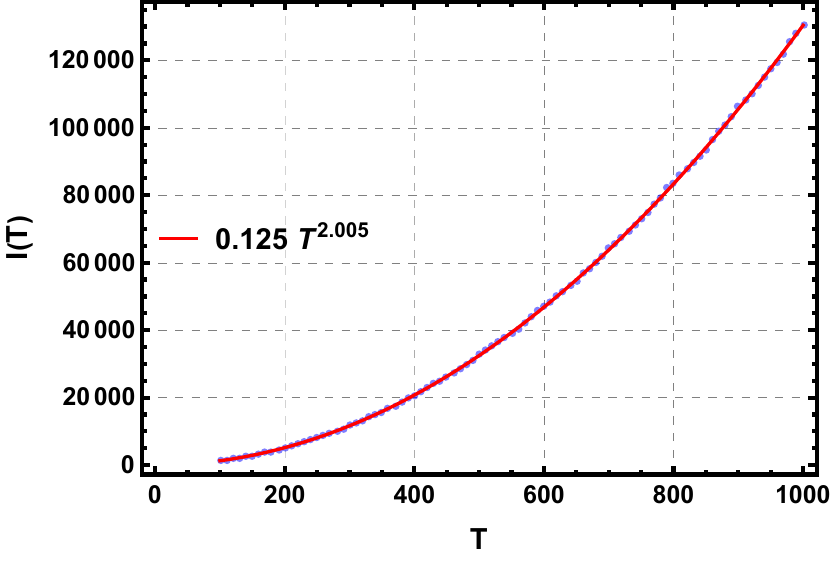}
    \end{subfigure}
    \caption{These figures display the behavior of $I(T)$ as a function of $T$ for two cases: $\beta = 0.1$ (left) and $\beta = 0$ (right). The fitting functions, obtained using Mathematica's \texttt{FindFit} command, show good agreement with the analytical expression given by the right-hand side of equation \eqref{timeavg}. Note that $0.125 \approx \pi/24$.}
    \label{FIT2}
\end{figure}

For this range of $\beta$, no well-defined mean value theorem exists. However, we can relate  $\zeta(\beta+it)$ for $\beta<1/2$ to $\zeta(1-\beta-it)$, where $(1-\beta)>1/2$, and then apply the mean value theorem \eqref{mean2}. The relevant relation is \cite{Titchmarsh}
\begin{equation}
    \zeta(s) = \chi(s) \zeta(1 - s),
\end{equation}
where $s=\beta+it$, and 
\begin{equation}
    \chi(s) = 2^s \pi^{s-1} \sin\left(\frac{\pi s}{2}\right) \Gamma(1 - s).
\end{equation}
In the limit $t \gg 1$, we can approximate $\chi(s)$ using Stirling's formula as
\begin{equation}
    |\chi(\beta + it)| \approx \left(\frac{t}{2\pi}\right)^{1/2 - \beta} \left(1 + O\left(\frac{1}{|t|}\right)\right),
\end{equation}
which implies
\begin{equation}
    |\zeta(\beta + it)|^2 \approx \left(\frac{t}{2\pi}\right)^{1 - 2\beta} |\zeta(1 - \beta + it)|^2.
\end{equation}
Note that here, and in some of the following steps in these calculations, we can (and do) drop terms coming from the lower limits of integrals because the statement we are trying to prove is expected to hold in the limit of large $T$.
The integral appearing in the time-averaged SFF is then
\begin{align}\label{int1}
    I(T) &= \int_{1}^{T} dt \zeta(\beta+i t) \zeta(\beta- it) \nonumber \\
    & \approx \left(\frac{1}{2\pi}\right)^{1 - 2\beta} \int_{1}^{T} t^{1 - 2\beta} |\zeta(1 - \beta + it)|^2 \, dt. 
\end{align}
Now we have the formula for the integration by parts
\begin{equation}
    \int_{a}^b u(t) v'(t) dt =u(b)v(b)-u(a)v(a)-\int_{a}^{b} u'(t)v(t) dt
\end{equation}
we choose $v(t)=\int_{1}^{t} |\zeta(1 - \beta + ir)|^2 dr$ with $t \gg 1$.
Applying this formula to \eqref{int1}, we obtain 
\begin{align}\label{intxx}
    \int_{1}^{T} t^{1 - 2\beta} |\zeta(1 - \beta + it)|^2\, dt 
    &= \left( t^{1 - 2\beta} \int_{1}^{t} |\zeta(1 - \beta + ir)|^2\, dr \right) \Big|_{1}^{T} \nonumber \\
    &\quad - (1 - 2\beta) \int_{1}^{T} t^{-2\beta} \left( \int_{1}^{t} |\zeta(1 - \beta + ir)|^2\, dr \right) dt \nonumber \\
    &= T^{1 - 2\beta} \int_{1}^{T} |\zeta(1 - \beta + ir)|^2\, dr \nonumber \\
    &\quad - (1 - 2\beta) \int_{1}^{T} t^{-2\beta} \left( \int_{1}^{t} |\zeta(1 - \beta + ir)|^2\, dr \right) dt.
\end{align}
Now, in the second term, for $t \gg 1$, we can approximate the inner integral as
\begin{equation}
    \int_{1}^{t} dr  |\zeta(1 - \beta + ir)|^2= \zeta(2-2\beta)t+O(t^{2\beta}).
\end{equation}
Substituting this into \eqref{intxx}, we get
\begin{align}
    \int_{1}^{T} t^{1 - 2\beta} |\zeta(1 - \beta + it)|^2 \, dt &=T^{2 - 2\beta} \zeta(2-2\beta) -(1-2\beta)\int_{1}^{T} dt \, t^{1-2\beta} \zeta(2-2\beta)+O(T) \nonumber \\
    &=T^{2 - 2\beta} \zeta(2-2\beta)-\zeta(2-2\beta) \frac{1-2\beta}{2-2\beta} T^{2-2\beta} +O(T) \nonumber \\
    &= \frac{\zeta(2-2\beta)}{2-2\beta} T^{2-2\beta}+O(T).
\end{align}
Therefore, we finally obtain
\begin{equation}\label{timeavg}
    I(T)= \int_{1}^T dt \zeta(\beta+i t) \zeta(\beta- it)\approx\left(\frac{1}{2\pi}\right)^{1-2\beta} \frac{\zeta(2-2\beta)}{2-2\beta} T^{2-2\beta}.
\end{equation}
For $\beta=0$, this simplifies to
\begin{equation}\label{sigma0}
    I(T)=\lim_{T\rightarrow \infty} \int_{1}^T dt \zeta(i t) \zeta(- it)= \frac{\zeta(2)}{4\pi} T^{2}=\frac{\pi}{24}T^2.
\end{equation}
The time-averaged SFF grows quadratically with $T$, implying that $g(t,0)$ is linear in $t$ (modulo fluctuations),
\begin{equation}\label{sfffit}
    \langle g(t, 0) \rangle \equiv \langle \zeta(i t) \zeta(- it) \rangle= \frac{\pi}{24} t.
\end{equation}
Figure \ref{FIT2} shows the plot of $I(T)$ as a function of $T$, demonstrating good agreement with the analytical result.

\begin{figure}
    \centering
    \includegraphics[width=.55\textwidth]{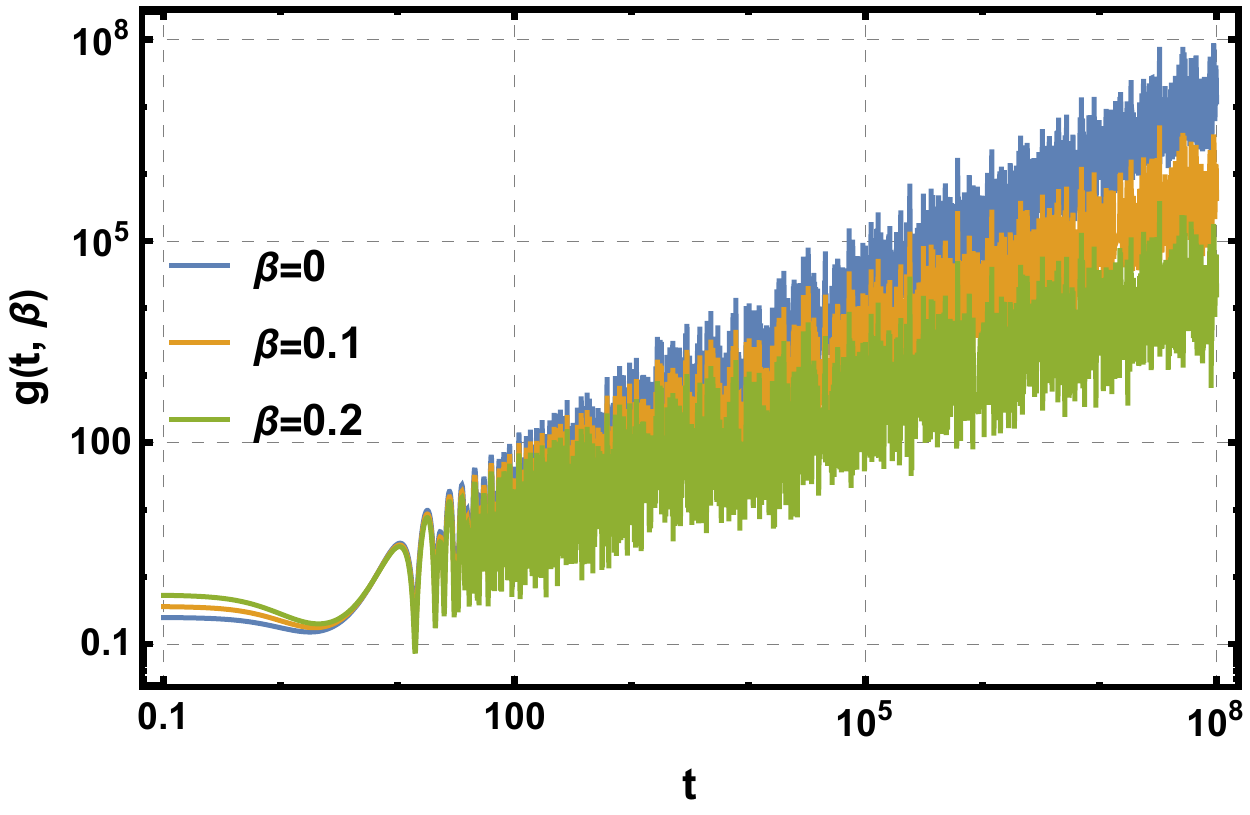}
   \caption{Behavior of $g(t, \beta)$ for various $\beta$, illustrating that the ramp becomes steeper as $\beta$ decreases, in agreement with equation \eqref{timeavg}.}
    \label{sff_comp}
\end{figure}

Figure \eqref{sff_comp} presents the SFF for various values of $\beta$, illustrating that the ramp becomes steeper as $\beta$ decreases, consistent with the behavior predicted by equation \eqref{timeavg}.

\newpage

\section{Thouless Time of the Log Spectrum}
A notable fact that follows from the above observations is that we can associate a Thouless time to the $\log n$ spectrum. It is generally believed \cite{ShenkerBPS} that a Thouless time that scales extensively with the system size corresponds to a weakly chaotic system, whereas an $\mathcal{O}(1)$ Thouless time is an indicator of strong chaos. Since the results of \cite{FirstRamp, FuzzRandom, SimplestRamp} show that the $\log n$ spectrum has very close connections to black hole normal modes (including various random matrix signatures), it will be interesting if its Thouless time also reflects black hole expectations. Remarkably, this turns out to be the case.

Black holes are expected to be strongly chaotic and therefore are expected to have an $\mathcal{O}(1)$ Thouless time. However, as far as we are aware, the only black hole-inspired systems which exhibit demonstrably $\mathcal{O}(1)$ Thouless times are random matrices themselves. No known black hole toy model in the bulk or boundary has been able to reproduce this explicitly. SYK model, which is a boundary toy model for black holes, has a Thouless time that scales as $\mathcal{O}(\log N)$ whereas bulk fuzzball solutions in supergravity scale as $\mathcal{O}(N^\#)$. It is believed that black hole microstates in $\mathcal{N}=4$ SYM should have an $\mathcal{O}(1)$ Thouless time, but it is not clear how to demonstrate this directly (from the bulk or boundary).

In earlier work \cite{FirstRamp, FuzzRandom, SimplestRamp, Burman1, Pradipta1, Burman2}, we obtained normal modes of black holes by postulating that microstates have no interior, and by computing normal modes for probe fields with non-dissipative boundary conditions at a Planckian stretched horizon. It was demonstrated that both thermodynamics and (exterior) correlator expectations about smoothness can be reproduced in this way. Furthermore the linear ramp and level repulsion could also be obtained, despite the fact that the system under consideration had nothing directly to do with random matrices. This is the context that makes the Thouless time of the $\log n$ spectrum interesting.

\begin{figure}
    \centering
    \includegraphics[width=.55\textwidth]{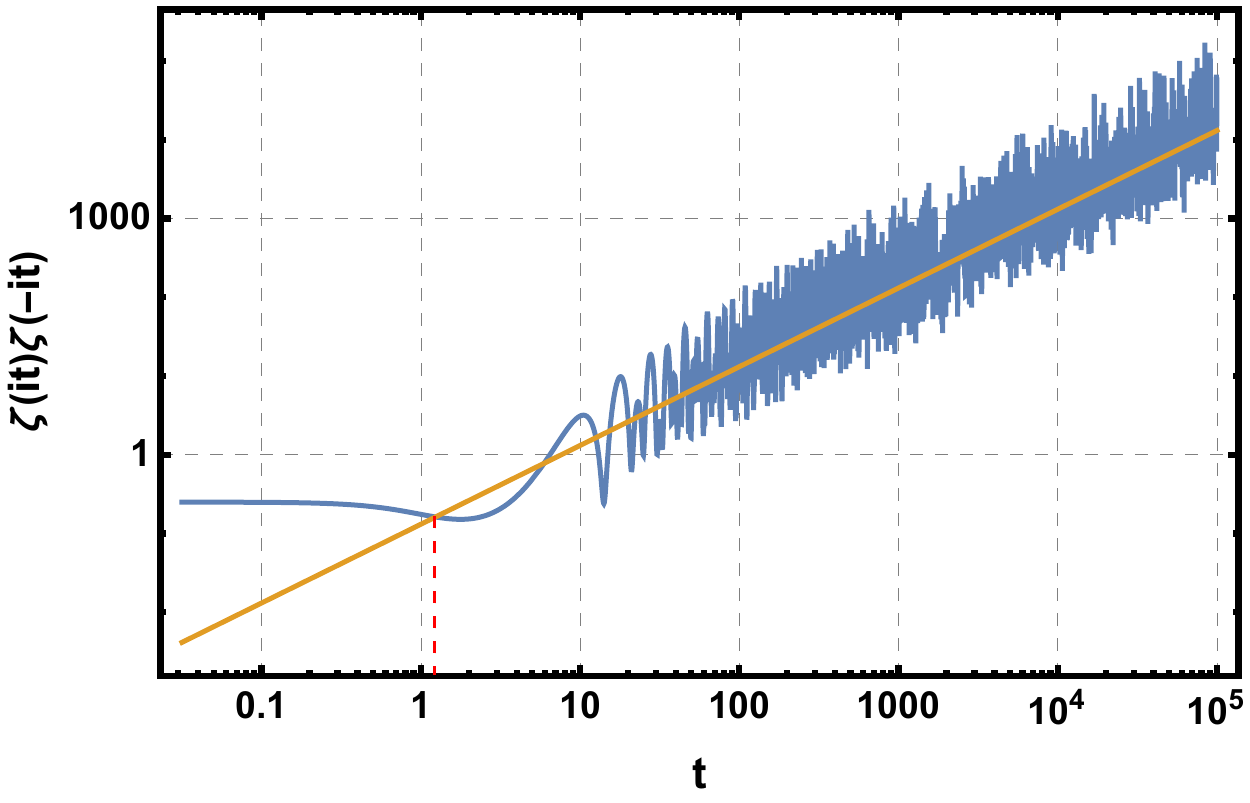}
   \caption{ This figure illustrates that the Thouless time of the log spectrum is $O(1)$. The blue curve shows the SFF at $\beta=0$ after subtracting the classical contribution, i.e., \( \zeta(it) \zeta(-it) \), while the yellow line represents the time-averaged slope $\frac{\pi}{12}t$. The red dashed line marks their first intersection, which is a measure of the Thouless time.}
    \label{thouless_log}
\end{figure}

Thouless time is usually extracted in RMT systems by studying the linear ramp, and by identifying the time at which the linear ramp starts \cite{HanadaShenker, ShenkerBPS}. However, there is a subtlety -- the ramp is a quantum effect reliant to the discreteness of the spectrum, and the ``true'' start of the ramp is often hidden by the classical dip. So in order to determine the Thouless time and the true beginning of the ramp, one often resorts to various tricks. These include techniques like Gaussian filtering \cite{HanadaShenker} or working with the connected unfolded SFF \cite{Dario}. The goal of all these operations is the same -- we wish to get rid of the effects of the dip without doing too much damage to the SFF so that the behavior of the quantum ramp can be reliably extracted. 

The remarkable fact is that the arguments in our previous sections show that the Riemann zeta function is {\em naturally} understood as the quantum contribution to the (complexified) partition function of the $\log n$ spectrum, truncated at integer $N$. Here $N$ should be viewed as the size of the system. It is a trivial matter to determine the ``beginning'' time of the full ramp from the plots of the mod square RZF as well our analytic average ramp function -- we do that in Figure \ref{thouless_log}. It should also be evident from (say Figure \ref{ramp_fit} ) that the RZF ramp is indeed capturing the ramp that was ``hidden under the dip''. It is immediate that the Thouless time is unambiguously $\mathcal{O}(1)$ from Figure \ref{thouless_log}. It is also naturally independent of $N$.

Our original motivation for studying the $\log n$ spectrum was its connection to black hole normal modes. The precise relationship between these normal modes and dynamical chaos may require a more complete understanding of UV complete black hole microstates in the bulk. But we view the fact that the linear ramp \cite{FirstRamp}, level repulsion \cite{FuzzRandom},  and now the correct Thouless time can all be obtained very simply from this landscape of ideas, encouraging.

\section{$L$-Functions}  
\label{sec:dirichlet_sff}  

We begin with the main conclusion of this section: our results for the zeta function have natural generalizations to more general (so-called) $L$-functions. 

$L$-functions are complex-valued functions that encode arithmetic and analytic properties of number fields, modular forms, and automorphic representations \cite{murty2024lfunction,peakmath_channel}. They generalize the Riemann zeta function and play a central role in mathematics. There are large classes of explicitly known {\em arithmetic} $L$-functions (see examples in Appendix \ref{app:zoo}), but their general analytic properties are conjectural. A general $L$-function can be viewed a a series (``Dirichlet series'') \cite{IwaniecKowalski,MontgomeryVaughan}
\begin{equation}  
L(s) = \sum_{n=1}^{\infty} \frac{a_n}{n^s}, 
\end{equation}  
but not all sequences $\{a_n\}$ yield $L$-functions. Selberg's criteria, which we summarize in Appendix \ref{app:selberg}, are believed to impose necessary analytic conditions on $\{a_n\}$ for $L(s)$ to be an $L$-function. Apart from certain boundedness and convergence properties, these criteria require that the function $L(s)$ have an analytic continuation into the complex $s$-plane, have a fairly mild singularity structure at $s=1$, satisfy a certain functional equation, and have an Euler product representation. The functional equation is what we will find to be crucial, in controlling the ramp.

The simplest case, with \( a_n = 1 \), corresponds to the Riemann zeta function. Another fundamental example is the Dirichlet $L$-function. We present some of the well-known classes of arithmetic $L$-functions in Appendix \ref{app:zoo}. For illustration, we present two explicit examples below:
\begin{itemize}
    \item The Dirichlet $L$-function associated with the Dirichlet character\footnote{The precise definitions of some of the number theoretic terms that we use here may not be familiar to physicists. They are easily looked up--but since our work relies more directly on the analytic side of the story, we will often not emphasize the number theoretic aspects. The subject is huge, so our approach here is pragmatic and we will only include those definitions that are essential for our purposes.} $\chi_{3,2}$:
$$L(s, \chi_{3,2}) = \frac{1}{1^s} - \frac{1}{2^s} + \frac{1}{4^s} - \frac{1}{5^s} + \frac{1}{7^s} - \frac{1}{8^s} + \frac{1}{10^s} - \frac{1}{11^s} + \ldots$$ Here $a_n \equiv n (\textrm{mod} 3)$.
\item For the elliptic curve $E: y^2 = x^3 - x$ with conductor\footnote{We will define what a conductor is, below.} 32:
$$L(E, s) = \frac{1}{1^s} - \frac{1}{3^s} + \frac{3}{5^s} - \frac{1}{7^s} + \frac{1}{9^s} - \frac{3}{11^s}  -\frac{3}{15^s} + \ldots$$
In the series above, $a_p = p + 1 - \#E(F_p)$ for a prime $p$, with $\#E(F_p)$ the number of points on the curve $E$ over the finite field $F_p$ (i.e., $F_p$ is the set consisting of the numbers $0$ through $p-1$, with addition and multiplication defined modulo $p$). See Appendix \ref{app:zoo} for some more detail. Other $a_n$'s are defined through multiplicity, e.g. $a_{15}=a_3 a_5$.
\end{itemize}

These explicit formulas show how the arithmetic information (character values, point-counting on the elliptic curve, ...) directly determines the non-zero coefficients in the series expansion of an arithmetic $L$-function.

\subsection{The Euler Product and Connection to Physics}
A remarkable feature of $L$-functions is that when the sequence $\{a_n\}$ satisfies appropriate arithmetic conditions, the $L$-function can be expressed as an Euler product \cite{murty2024lfunction}:
\begin{equation}
L(s) = \prod_{p \in\text{primes}} f_p(p^{-s})
\end{equation}
where $f_p$ is typically a rational function. For the Riemann zeta function, this takes the simple form (look at Appendix \ref{AppSecQuant}):
\begin{equation}
\zeta(s) = \prod_{p \in\text{primes}} \frac{1}{1-p^{-s}}
\end{equation}
For the Dirichlet $L$-function and elliptic curve $L$-function examples we discussed above, the Euler products are:
\begin{equation}
L(s, \chi_{3,2}) = \prod_{p \in\text{primes}} \frac{1}{1-\chi_{3,2}(p)p^{-s}}
\end{equation}
\begin{equation}
L(E, s) = \prod_{p \nmid 32} \frac{1}{1-a_p p^{-s} + p^{1-2s}}
\end{equation}

From a physicist's perspective, the Euler product representation bears a striking resemblance to the second quantized\footnote{We describe the second quantized perspective in detail, for the Riemann zeta function, in an Appendix.} grand canonical partition function in statistical mechanics:
\begin{equation}
\mathcal{Z} = \prod_i \left(1 + e^{-\beta(E_i-\mu)}\right)
\end{equation}
The parameter $s$ in the $L$-function plays a role analogous to inverse temperature $\beta$, while the logarithm of prime numbers can be interpreted as first quantized energy levels. The coefficients $a_n$ encode information analogous to the chemical potential $\mu$, determining the statistical weights of different number-theoretic states. Just as the chemical potential controls the average particle number in a thermodynamic system, the arithmetic coefficients $\{a_n\}$ control the number-theoretic properties of the $L$-function.

This connection suggests that the zeros of $L$-functions might be understood as phase transitions in a corresponding statistical mechanical system, where the analytical properties reveal critical phenomena in the underlying number-theoretic structure. The Riemann Hypothesis and its generalizations to other $L$-functions could then be interpreted as statements about the universality of these phase transitions, providing possible physical intuition for these mathematical conjectures. 

\subsection{Functional Equation and Degree of an $L$-function:}
A defining property of an \( L \)-function is the existence of a functional equation. For a general $L$-function, the functional equation takes the form\footnote{We follow the notations of \cite{lmfdb}.}:  
\begin{equation}  
\Lambda(s) \equiv N^{s/2}  
\prod_{j=1}^{J} \Gamma_{\mathbb{R}}(s+\mu_j) \prod_{k=1}^{K} \Gamma_{\mathbb{C}}(s+\nu_k)  
\cdot L(s+w/2) = \varepsilon \overline{\Lambda}(1-s), \label{LMFDB-FunEq}
\end{equation}  
where $w$ is a positive integer called weight\footnote{The number $\frac{w+1}{2}$ is the real part of the critical line of the $L$-function. The modified grand Riemann hypothesis is the assertion that the nontrivial zeros of all automorphic $L$-functions lie on the critical line.} and \( N \) is a positive integer called conductor. \( \Gamma_{\mathbb{R}} \) and \( \Gamma_{\mathbb{C}} \) are gamma function terms defined as:
\[
\Gamma_\mathbb{R}(s) := \pi^{-s/2}\Gamma(s/2)\qquad\text{and}\qquad \Gamma_\mathbb{C}(s):= 2(2\pi)^{-s}\Gamma(s)
\]
The integer  
\begin{equation}  
d = J + 2K  
\end{equation}  
is called the degree, and the complex number $\varepsilon$ has unit modulus, and it is called the sign. Collectively, the set of parameters $(d, N,(\mu_1,\ldots,\mu_J:\nu_1,\ldots,\nu_K),\varepsilon)$  form the Selberg data\footnote{For a more detailed introduction to the parameters, please refer to LMFDB \cite{lmfdb}.}. The notation $\overline{\Lambda} (z)$ means that the parameters of the function get complex conjugated, but its argument $z$ does not. In effect, this means that $\overline \Lambda(z) \equiv \overline{\Lambda(\overline z)}$. For known arithmetic $L$-functions, the $\mu_j$ and $\nu_k$ (which are collectively referred to as spectral parameters) are real. This is important in our discussion of the ramp.

The degree of an $L$-function is a fundamental invariant measuring its complexity. We will soon discuss in the next section that the ramp-plateau behavior of an $L$-function is governed by its degree. The Riemann zeta function and Dirichlet $L$-functions have degree $d=1$, reflecting their connection to $\mathbb{Q}$ and Dirichlet characters respectively. For a modular form $f$ of weight $k$, the associated $L$-function $L(f,s)$ has degree $d=2$, with the weight $k$ instead influencing the $\Gamma$-factors in the functional equation. In particular, the $L$-function of an elliptic curve has degree $2$, consistent with the modularity theorem that relates it to a weight $2$ modular form. More generally, higher-degree $L$-functions appear in the study of higher-dimensional arithmetic objects and automorphic representations. 

\subsection{Results}
We now highlight some key results and discuss their relevance for general \( L \)-functions, building on what we have already seen for the Riemann \( \zeta \)-function.

\subsubsection{Hagedorn Transition} There is a pole at \( s = 1 \) for the Riemann \( \zeta \)-function, which we interpreted as a Hagedorn transition. The pole at $s=1$ is not generally present in other \( L \)-functions, but the location $s=1$ is often special.
    
    \subsubsection{Ramp-Plateau Structure} For \( \beta < \frac{w+1}{2} \), the characteristic ramp-plateau structure persists across a broad class of \( L \)-functions. As with the Riemann \( \zeta \)-function, the truncated version displays both ramp and plateau behavior (see Appendix \ref{truncated_L}), while the non-truncated version exhibits an infinite ramp.  We deduce that at $\beta=0$ the exponent of the ramp is given by twice the product of its degree  and the real part of the central line. 
We derive these statements using results in the literature for \( L \)-functions \cite{IwaniecKowalski}. 

\begin{figure}
    \centering
    \includegraphics[width=.55\textwidth]{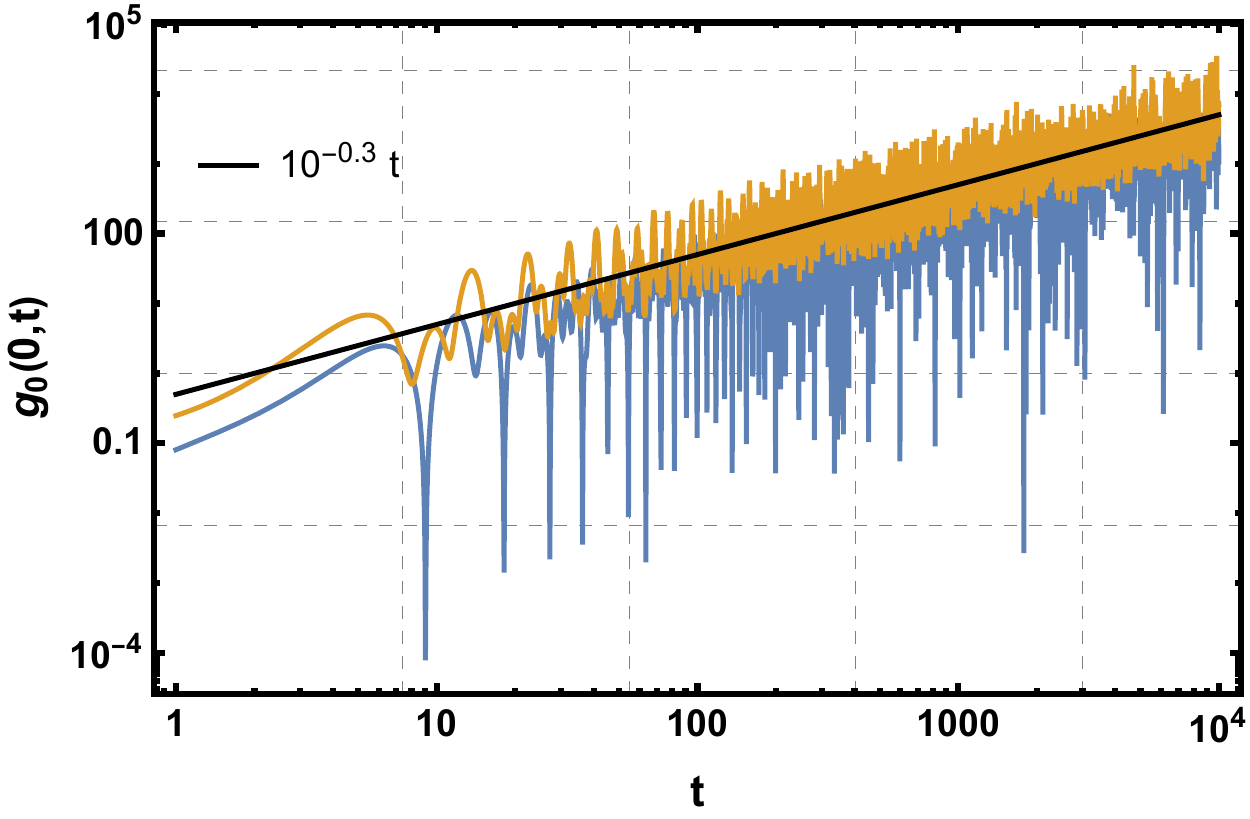}
   \caption{SFF at $\beta = 0$ constructed from two different Dirichlet $L$-functions (defined in \ref{DLfn}): \texttt{DirichletL[2,1,0 + it]} (blue) and \texttt{DirichletL[3,2,0 + it]} (yellow). Both curves exhibit an eternal ramp, similar to the behavior observed in the Riemann zeta function. A black line with unit slope is overlaid to indicate that the ramp has slope one.}
    \label{comparison_DirichletL}
\end{figure}

\begin{figure}
    \centering
    \includegraphics[width=.75\textwidth]{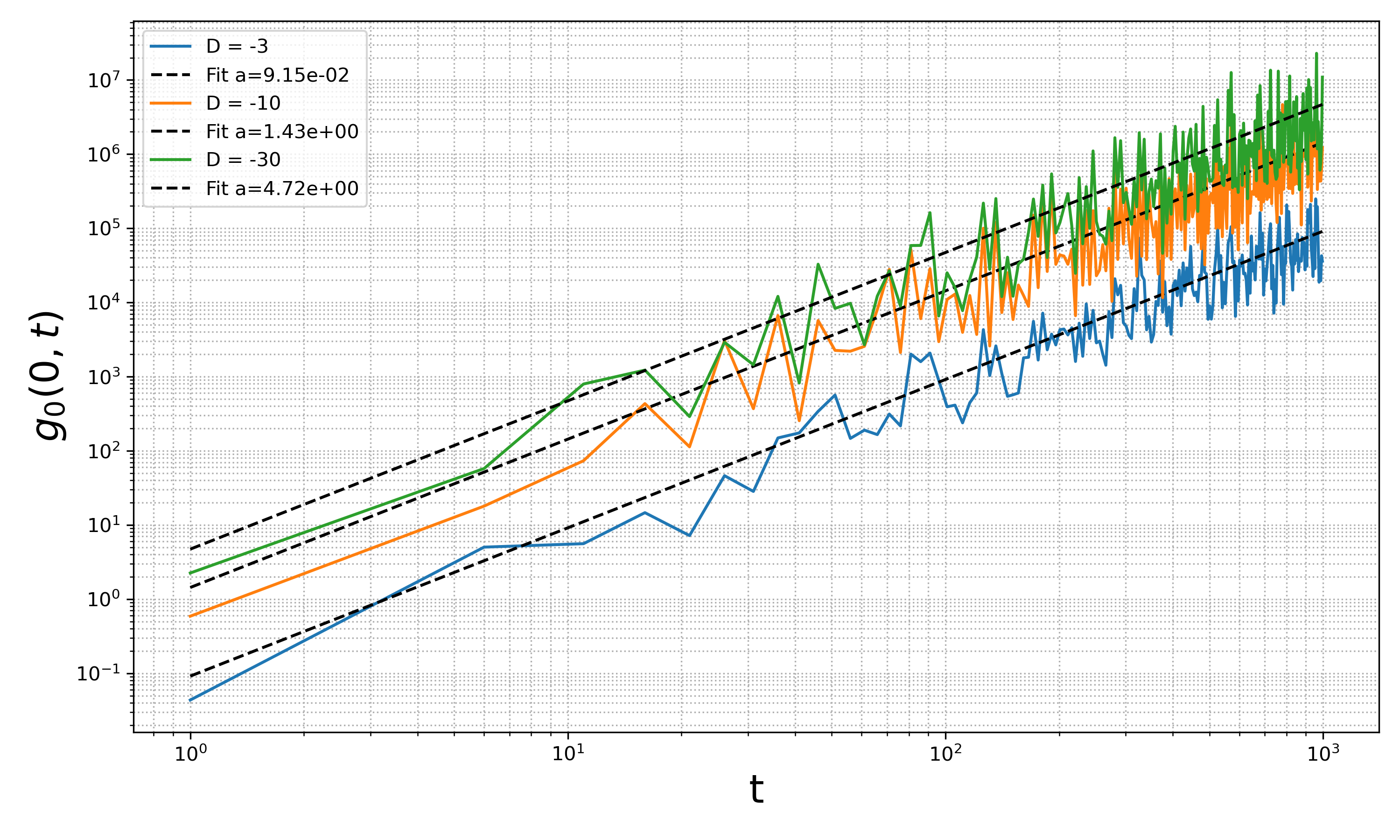}
   \caption{SFF at $\beta = 0$ constructed from Dedekind zeta functions (as defined in Eq.~\eqref{DeDefn}) corresponding to three different quadratic fields $\mathbb{Q}(\sqrt{D})$. The plots exhibit an eternal ramp that is well approximated by a quadratic fit of the form $y = a t^2$, with the fitted values of $a$ indicated in the figure. This suggests a quadratic growth behavior in the ramp region. }
    \label{comparison_Dedekind}
\end{figure}

\begin{figure}
    \centering
    \includegraphics[width=.75\textwidth]{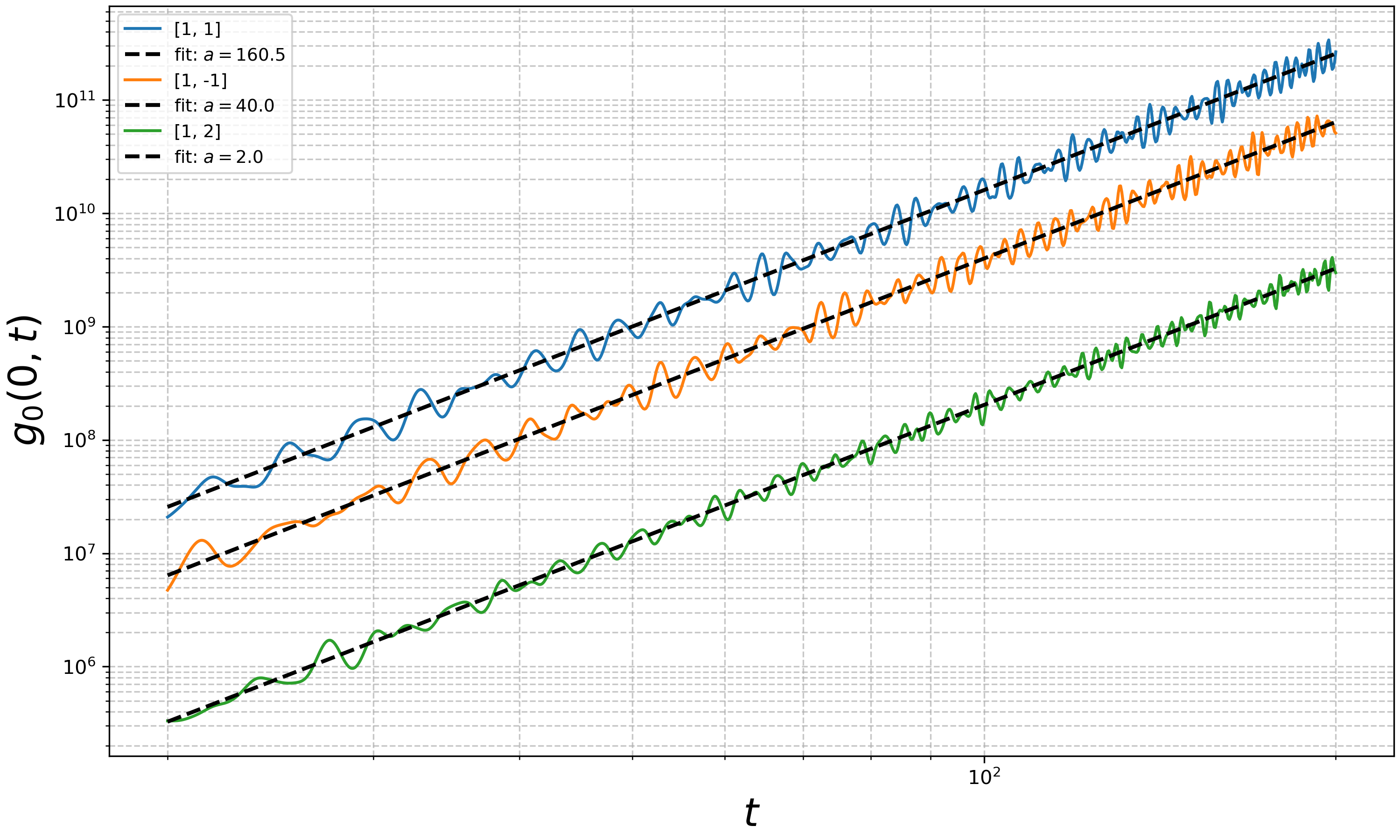}
   \caption{Log-log plot of the Spectral Form Factor (SFF) for \( L \)-functions of elliptic curves defined by \( y = x^3 + \mu x + \nu \), with the parameters \( [\mu, \nu] \) chosen from three different sets. The SFF is fitted to a quartic function of the form \( y = at^4 \), with the fitted parameter values shown in the figure. This suggests that the \( L \)-functions exhibit an eternal ramp with quartic growth.}
    \label{comparison_elliptic}
\end{figure}

Here, we directly work with the untruncated spectral form factor (SFF) as:
\begin{equation}
    g_0(\beta, t) = L(\beta + it, \chi_j) L(\beta - it, \chi_j).
\end{equation}
For a wide class of \( L \)-functions (including Dirichlet and automorphic \( L \)-functions), we observe the following:

\begin{itemize} 
\item \textbf{\( \beta > \frac{w+1}{2} \)}: The second moment on the imaginary axis is a constant \( c_1 \) \cite{IwaniecKowalski}:
\begin{equation}
     \frac{1}{T} \int_0^T g_0(\beta, t) \, dt \to c_1, \quad \text{as} \quad T \to \infty.
\end{equation}

\item \textbf{\( \beta = \frac{w+1}{2} \)}: This is the critical line. It will not be our focus here. However the literature suggests that for many $L$-functions (including zeta, Elliptic and Dirichlet-$L$) \cite{Titchmarsh,IwaniecKowalski}, the second moment on the critical line grows asymptotically as
\begin{equation}
\int_0^T g_0\left(\frac{w+1}{2}, t\right) \, dt \sim c_2 \cdot T \log T, \quad \text{as} \quad T \to \infty,
\end{equation}
where the constant \( c_2 \) depends on the specific \( L \)-function.

\item \textbf{\( \beta < \frac{w+1}{2} \)}:
Here we expect a ramp. To find out the exponent,
we begin with the functional equation:  
\begin{align}
\Lambda(\beta + it) \sim \overline{\Lambda}(1 - \beta - it),
\end{align}
where we have suppressed the sign because it drops out in our absolute value expressions. This yields the following expression for \( L(\beta + it) \):  \begin{align*}  
&L(\beta + it+w/2) \sim \\
&N^{\frac{1 - 2\beta}{2}} 
\frac{\prod_{j=1}^{J} \Gamma_{\mathbb{R}}(1 - \beta - it + \mu_j)}{\prod_{j=1}^{J} \Gamma_{\mathbb{R}}(\beta + it + \mu_j)} 
\frac{\prod_{k=1}^{K} \Gamma_{\mathbb{C}}(1 -\beta - it + \nu_k)}{\prod_{k=1}^{K} \Gamma_{\mathbb{C}}(\beta + it + \nu_k)} \overline{L}(1 - w/2-\beta - it).
\end{align*}
Assuming that \( \overline{L}(1 -w/2- \beta - it) = \mathcal{O}(1) \)\cite{MontgomeryVaughan,IwaniecKowalski}, and applying Stirling's formula to the gamma ratios (as referenced in Equation~\ref{GammaRatio})\footnote{Note that the reality of spectral parameters was important in this step.}, we obtain—up to a constant factor and a pure phase—the asymptotic behavior:
\begin{align}
|L(\beta + w/2+it)| &\sim t^{(J/2 + K)(1 - 2\beta)}, \\
\Rightarrow \quad g_0(\beta, t) &\sim t^{(J + 2K)(1 - 2\beta+w)}, \\
\Rightarrow \quad g_0(0, t) &\sim t^{(J + 2K)(1+w)} = t^{2 d \frac{w+1}{2}},
\end{align}
Hence we have ramp exponent=$2 \times d \times $ real part of the central line 
where \( d = J + 2K \) denotes the degree of the \( L \)-function.
Our main focus will  \( \beta = 0 \). We present numerical evidence supporting our analytic conclusion in Figure \ref{comparison_DirichletL}, \ref{comparison_Dedekind}, \ref{comparison_elliptic}.

\end{itemize}

%

\section{Origin of Ramp in Other Sequences}\label{sec:OtherSeq}

In this section, we will analytically demonstrate that the slope of the ramp in the SFF constructed from $E_n=\{n^{\alpha} \}$ is indeed $1/(1-\alpha)$, as suspected in \cite{SimplestRamp}, by using the Poisson resummation formula.

Poisson resummation formula is a fundamental result that connects sums of a function over integers to sums of its Fourier transform over frequencies. At its core, it nothing more than a change of basis, switching between the integer lattice and its dual frequency lattice. But this simple idea links number theory, harmonic analysis, and statistical physics and it is conceivable that it is at the root of many dualities in physics. 

The Poisson resummation formula is given by:
\[
\sum_{n \in \mathbb{Z}} f(n) = \sum_{k \in \mathbb{Z}} \hat{f}(k),
\]
where \( \hat{f}(\xi) \) is the Fourier transform of \( f \), defined as:
\[
\hat{f}(\xi) = \int_{-\infty}^\infty f(x) e^{-2\pi i \xi x} \, dx.
\]
We will apply the above formula to  study our ramp functions.

Let us begin with an arbitrary partition function:
\begin{equation}
    Z(s) = \sum_{n=0}^{N} \exp\left(-s w(n)\right),
\end{equation}
where \( w(n) \) is a set of increasing numbers representing the eigen frequencies or energy levels of the system.  For example, we may choose \( w(n) = \sqrt{n} \). The ultraviolet cutoff $N$ may (or may not) go to infinity.  In most physical situations, \( w(n) \) is bounded from below, and we will assume this to be the case, and without loss of generality we will assume $w(0)=0$.

We will take $f(x)=\exp(-s w(x))$ when $0 \le x \le N$ and zero elsewhere. We get,
\[
Z(s)=\sum_{n \geq 0}^N e^{-s w(n)} =   \sum_{k \in \mathbb{Z}}  \hat{f}(k)=Z_{cl}(s)+ \sum_{k \in \mathbb{Z+}} (\hat{f}(k)+\hat{f}(-k)),
\]
with,
\[
\hat{f}(k)=\int_0^N e^{-s w(x)} e^{-2\pi i k x} dx.
\]
We have defined 
\[ Z_{cl(s)}=\hat{f}(0)=\int_0^N e^{-s w(x)} dx.\] The zero-mode integral can be thought of as the classical limit of the partition function when the frequencies are continuous. The Ramp function is simply defined as (this is precisely analogous to our RZF discussion),
\begin{align}\label{RampFunction}
  R(s)=Z(s)-Z_{cl}(s)= \sum_{k \in \mathbb{Z+}} (\hat{f}(k)+\hat{f}(-k)) 
\end{align}
For large-$s$ we can evaluate the $\hat{f}(k)$'s by saddle point method to get an approximation of the partition function.
\begin{eqnarray}
    -s w'(x_c)-2 \pi i k= 0, \   \implies \
    \hat{f}(k) \approx \sqrt{\frac{2\pi}{s w''(x_c)}} e^{-s w(x_c)-2 \pi i k x_c}  \label{saddle}
\end{eqnarray}
The approach is easiest illustrated with an example.

\subsection{Example: $w(n)=\sqrt{n}, s=i t, \beta=0, N=\infty$}

\begin{figure}[h]
    \centering
    \includegraphics[width=0.6\textwidth]{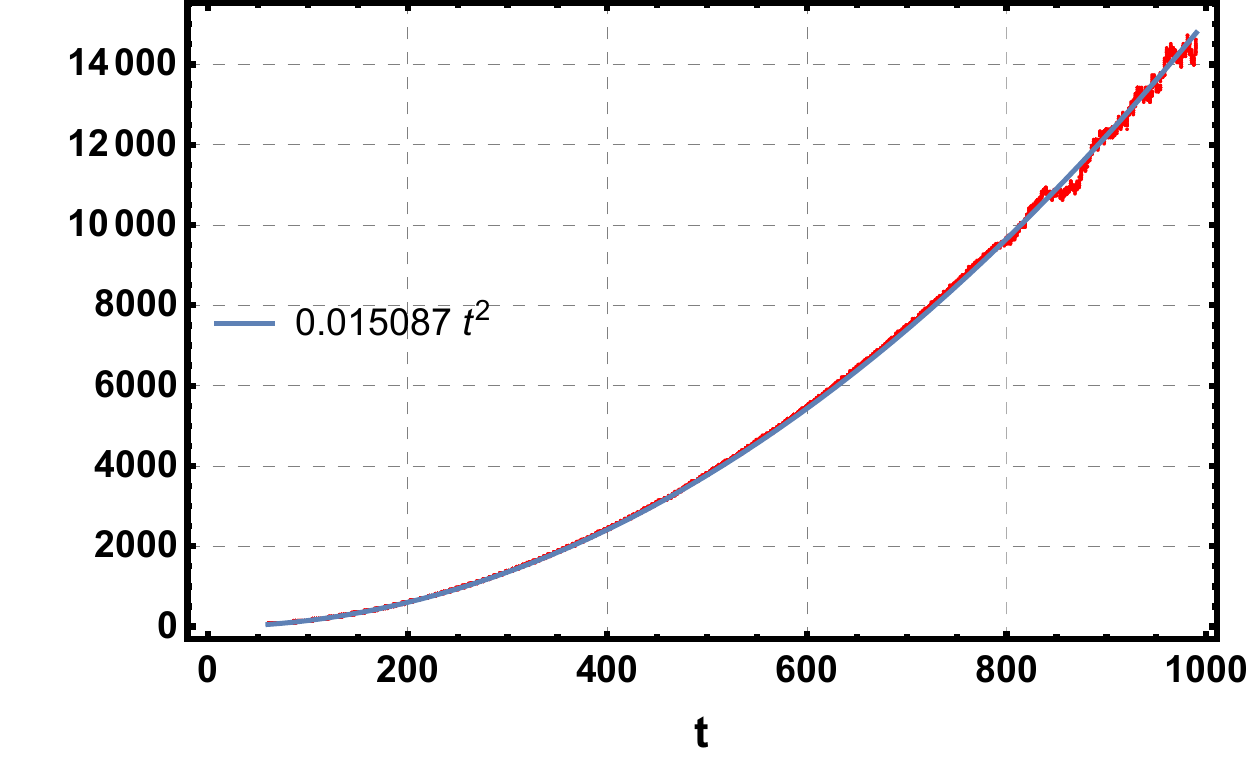}
    \caption{Smoothed ramp region of the spectral form factor (SFF) using a running average (\textit{Mathematica}'s \texttt{MovingAverage}, window size 200). The SFF is computed from energy levels \( E_n = \{\sqrt{n}\} \) with \( N = 10^4 \), showing the expected quadratic growth of the ramp.}
    \label{root_n}
\end{figure}

As the cutoff $N=\infty$, we may expect an infinite ramp here. Note that because we are working with real time, now the saddle is at late times. The saddle point equation is
\begin{align}
    -\frac{ t}{2 \sqrt{x_c}}= 2 \pi k.
\end{align} 
As by definition $\Re(\sqrt{x})>0$, solution exists only for $k < 0$. We have 
\begin{eqnarray}
\hat{f}(k)\approx \frac{\sqrt{i} \, t}{2\sqrt{2}\pi |k|^{3/2}}
 e^{\frac{i t^2}{8 k \pi}}
\end{eqnarray}
where $k$ is understood to be $ <0$. 
Hence 
\begin{align} 
R(i t)  \approx \sqrt{i} \frac{t}{2\sqrt{2} \pi} \sum_{k > 0}  |k|^{-3/2}  e^{\frac{-i t^2}{8 k \pi}}, \end{align}
%
If we calculate something like a mean value, it would behave as  
\begin{align} \label{MRT}
    \langle R(it) R(-i t) \rangle &= \frac{1}{T} \int_0^T R(it) R^{*}(it) \, dt \nonumber \\
          &\sim \frac{\zeta(3)}{24 \pi^2} T^2.
\end{align}
In Figure~\ref{root_n}, we present a smoothed version of the ramp region of the spectral form factor (SFF), obtained using a running average with Mathematica's \texttt{MovingAverage} function and a window size of 200. The SFF is constructed from the energy levels \( E_n = \{\sqrt{n}\} \) with \( N = 10{,}000 \). As expected from Eq.~\eqref{MRT}, the ramp exhibits quadratic growth, \( \sim a t^2 \). The theoretical prediction \( \zeta(3) / 8\pi^2 = 0.0152242 \) is within the error bars of our best-fit value of \( a = 0.015087\)\footnote{The factor we are comparing contains 8 instead of 24 because we are plotting $R(it)R(-it)$ as implemented by \texttt{MovingAverage} in Mathematica,  not $\langle R(it)R(-it)\rangle$. To illustrate this with a simple example, consider the function $y(t)=t^2$. The definition of the time average above will give $\langle y(t) \rangle=t^2/3$, but a plot of $y(t)$  with Mathematica's \texttt{MovingAverage} will yield a coefficient of 1 instead of 1/3.}.

\subsection{Example: General $w(n)=n^{\alpha}$}

We will now analytically demonstrate that the slope of the ramp, constructed from the sequence $E_n=\{n^{\alpha} \}$ with $0<\alpha<1$, is given by $\frac{1}{1-\alpha}$ in a log-log plot, i.e., the ramp scales as $t^{\frac{1}{1-\alpha}}$. For $w(n)=n^{\alpha}$, $\hat{f}(k)$ can be evaluated from our saddle point approach to be (up to an $\alpha$-dependent pre-factor that we do not track explicitly)
\begin{equation}
    \hat{f}(k)\approx s^{\frac{1}{2(1-\alpha)}} |k|^{-\frac{2-\alpha}{2(1-\alpha)}} \,  \exp \left(s^{\frac{1}{1-\alpha}} k^{-\frac{\alpha}{1-\alpha}} \right).
\end{equation}
This leads to
\begin{equation}
    R(i t) \sim t^{\frac{1}{2(1-\alpha)}}  \sum_{k \neq 0} \frac{1}{|k|^{\frac{2-\alpha}{2(1-\alpha)}}} \exp\left((i t)^{\frac{1}{1-\alpha}} k^{-\frac{\alpha}{1-\alpha}} \right).
\end{equation}
Consequently
\begin{align}
    \langle R(iT)R(-i T) \rangle &= \frac{1}{T} \int_0^T R(it) R^{*}(it) \, dt \nonumber \\
    & = \# \,  T^{\frac{1}{1-\alpha}}.
\end{align}
This result explains the numerically observed slope of the ramp in \cite{SimplestRamp}. Note that since we were only interested in the slope and not the pre-factor, the saddle location and the second derivative $w''(x_c)$ in \eqref{saddle} were all that mattered. Most of the rest of the quantities in the above expressions did not really play a role.

\section{Discussions}

We have seen that the $\log n$ spectrum, truncated at an integer $N$, has various features that one might attribute to the eigenvalues of a quantum mechanical system that is strongly mixing, and is reminiscent of black holes. In previous work, we noted that this system has a linear ramp with fluctuations and (when there is a small amount of noise in the spectrum \cite{FuzzRandom}) conventional level repulsion. In the present paper, we further showed that there is a natural way in which we can associate a Thouless time to the $\log n$ spectrum. This Thouless time turned out to be $\mathcal{O}(1)$, which is again something we expect from black holes \cite{ShenkerBPS}. No other black hole toy model that we are aware of (besides of course random matrices themselves) has a Thouless time of $\mathcal{O}(1)$.

These facts are intriguing, and it will be very useful to understand them better. Features of the RZF played a crucial role in our calculations. It is known that level correlations of the non-trivial zeroes of the RZF have connections to random matrices \cite{dyson1962statistical}. Our observations in this paper were in many ways easier to study because they seem more directly related to the functional equation rather than the zeroes themselves. 

After discussing the $\log n$ spectrum, we generalized our approach in two different ways. The first generalization can be called ``mathematical'', in that we looked for number theoretic functions that go beyond the RZF. These were the $L$-functions. They have universal properties that are analogous to those of RZF. Using these specific properties, we were able to argue (and numerically demonstrate) the structure of ramps in these systems. It will be very interesting to understand these general $L$-functions in terms of a second quantized physical perspective. The second, more ``physical'' generalization, was to move away from the $\log n$ spectrum. We were able to develop methods based on Poisson resummation, to determine the ramp in some of these cases.

Apart from the potential significance for black hole microstates, our results may have implications for $p$-adic AdS/CFT \cite{Gubser_2017} where the $\zeta$ function emerges naturally. A second possibility worth investigating is the significance of early time chaos in these discussions. Our discussion in this paper was often reliant on the spacing between levels, and therefore is more naturally viewed as an analysis of late time physics. To investigate early time chaos, OTOCs are often used instead. Since we are working directly with the spectrum, and do not have a notion of ``simple operators'', it is difficult to formulate a correlator calculation. But it may be possible to do it by retracing the steps that lead us to the normal modes in the first place, and working with the (scalar) field in the black hole background with a stretched horizon. It will be interesting to see if one can see an analogue of the chaos bound \cite{Maldacena_2016} using black hole normal modes. The calculations in \cite{Burman1, Burman2} dealt (effectively) with time-ordered correlators, and therefore were only aware of more simple real-time thermal physics. 


\section{Acknolwedgments}

We thank Sam van Leuven for discussions, Pradipta Pathak for help with the material in Appendix \ref{AppA1}-\ref{AppA2} and Vishnu Jejjala for comments on an earlier version of the draft. We thank ChatGPT and DeepSeek for discussions as well as help with improvements in text.

\appendix

\section{First, Second and Third Quantizations of the Primon Gas }\label{AppSecQuant}

We start by pedagogically reviewing the connection between first and second quantizations for general systems. We also review how one can produce the partition function of a second quantized system from that of the first quantized one, via the Plethystic exponential. It is known that the log spectrum can be viewed as the second quantization of the so-called primon gas \cite{JuliaNumbers, BakasGases}. Putting these ingredients together, we will see that the RZF partition function can be obtained from a plethystic exponential on the primons\footnote{The material in the first two subsections of this Appendix was worked out in collaboration with Pradipta Pathak.}.

Our results in the main body of the paper do not rely too directly on the primon gas origins of the log spectrum. But since the Hagedorn temperature associated to the primon gas was a crucial bit of our intuition, we feel that the relationship to primons may be important. We therefore summarize some of the relevant facts in this Appendix.

We will conclude this Appendix by noting a potential connection between a suitable third quantization of the primon gas (or equivalently, a second quantization of the log spectrum) and black hole microstates. 

\subsection{First and Second Quantizations}\label{AppA1}

Let us consider a quantum mechanical system with the spectrum $E_n = \omega_n$ (bounded from below). The partition function of this system is given by
\begin{eqnarray}
Z_1= \sum_{n=0}^{\infty} e^{-\beta \omega_n}.
\end{eqnarray}
We will call this the first quantized partition function. 

Second quantization is the process by which we produce a new system whose Hilbert space is the Fock space obtained by multi-particling the first quantized states. This means that we are now considering a new Hamiltonian 
\begin{eqnarray}
H=\sum_{n=0}^{\infty}\omega_n \ b_n^\dagger b_n   
\end{eqnarray}
where $b_n, b_n^\dagger$ are annihilation-creation operators. The second quantized partition function is
\begin{eqnarray}
Z_2={\rm Tr}(e^{-\beta H}) 
\end{eqnarray}
where the trace is over the Fock space. More explicitly,
\begin{eqnarray}
Z_2=\langle E_{\{N_n\}}| e^{-\beta \sum_{n=0}^{\infty} \omega_n \ b_n^\dagger b_n} |E_{\{N_n\}}\rangle = \sum_{N_0,N_1,N_2,...=0}^{\infty} e^{-\beta \sum_{n=0}^{\infty} N_n \omega_n}.
\end{eqnarray}
The eigenstates of the second quantized Hamiltonian $H$ are labeled by lists of non-negative integers $\{N_n\}$ such that 
\begin{eqnarray}
E_{\{N_n\}}= \sum_{n=0}^{\infty} N_n \ \omega_n.   
\end{eqnarray}
We have used this in the previous expression. The partition function simplifies to
\begin{eqnarray}
  Z_2&=& \sum_{N_0,N_1,N_2,...=0}^{\infty} e^{-\beta \omega_0 N_0}  e^{-\beta \omega_1 N_1} ... = \left(\sum_{N_0=0}^{\infty} e^{-\beta \omega_0 N_0}\right) \times \left(\sum_{N_1=0}^{\infty} e^{-\beta \omega_1 N_1}\right) \times ... \nonumber 
\end{eqnarray}
where we have used the fact that $\sum_{n=0}^{\infty}a^n \times \sum_{m=0}^\infty b^m = \sum_{n,m=0}^{\infty} a^n b^m$. The final result is
\begin{eqnarray}  
  Z_2 = {\rm Tr}(e^{-\beta H})=\prod_{n=0}^{\infty}\frac{1}{1-e^{-\beta \omega_n}}.
\end{eqnarray}

The above structure is of course, entirely standard. The only reason we have been explicit in the formulas is because usually, they are worked out for the special case $\omega_n= n \omega_0$. This is the familiar case of the harmonic oscillator. Second quantization of the harmonic oscillator leads to free field theory and the Planckian spectrum. Our goal above was to emphasize that the structure is in fact entirely general, and not dependent on the modes being those of free linear oscillators.

\subsection{Plethystic Exponential}\label{AppA2}

We define the plethystic exponential via 
\begin{eqnarray}
    P (\exp(f(x_1, ... , x_n)) \equiv \exp\sum_{\ell=1}^{\infty}\left(\frac{1}{\ell}f(x_1^\ell,...,x_n^\ell)\right)
\end{eqnarray}
If we only have one variable, which we will call $x_1 \equiv e^{-\beta} \equiv q$, the first quantized partition function is
\begin{equation}
    Z_1=\sum_{n=0}^\infty q^{\omega_n}.
\end{equation}
The claim is that the plethystic exponential of this object is the second quantized partition function. This is easy to demonstrate:
\begin{eqnarray}
  P(\exp(Z_1)) &=& \exp\sum_{\ell=1}^{\infty} \frac{1}{\ell}\sum_{n=0}^{\infty}(q^\ell)^{\omega_n} =\exp\sum_{n=0}^{\infty}\sum_{\ell=1}^{\infty} \frac{(q^{\omega_n})^\ell }{\ell}=\prod_{n=0}^\infty\exp \sum_{\ell=1}^{\infty} \frac{(q^{\omega_n})^\ell }{\ell} \nonumber\\
  &=&\prod_{n=0}^\infty\exp \left[-\log (1-q^{\omega_n})\right]=\prod_{n=0}^{\infty}\frac{1}{1-e^{-\beta \omega_n}}.
\end{eqnarray}

\subsection{Primon Gas}\label{AppPrimon}

\begin{figure}
\begin{subfigure}{0.51\textwidth}
    \centering
    \includegraphics[width=\textwidth]{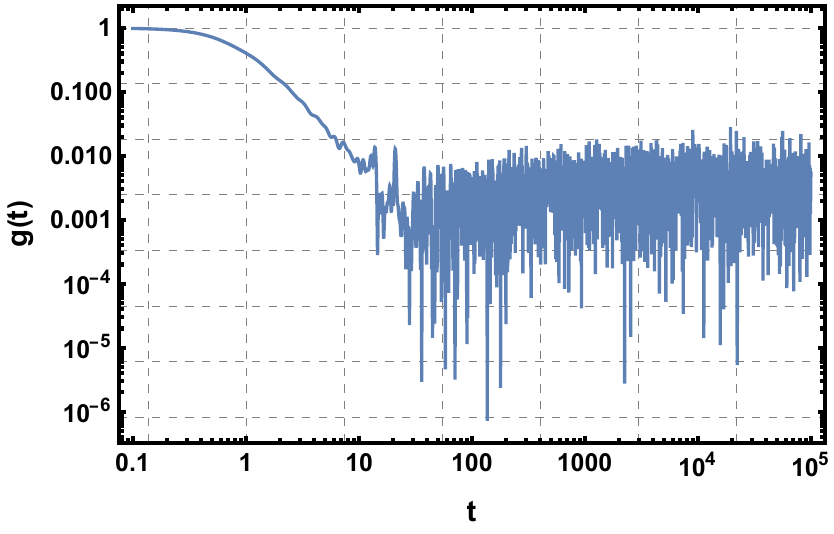}
    \end{subfigure}
    \hfill
    \begin{subfigure}{0.53\textwidth}
    \includegraphics[width=\textwidth]{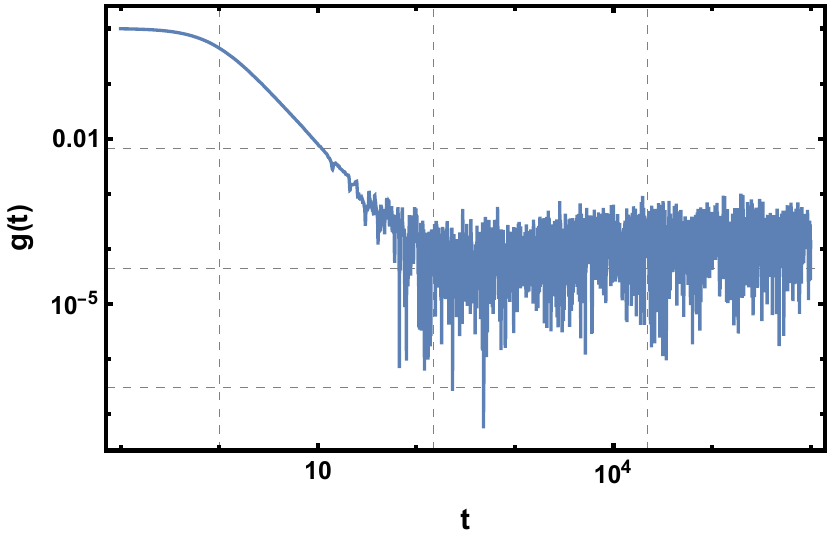}
    \end{subfigure}
    \caption{SFF for $\{E_n\} = \{\log p_n \}$ with $p_n$ as the $n$-th prime, shown for $p_{\text{max}} = 200$ (left) and $5000$ (right) at $\beta = 0$. A weak ramp appears at low $p_{\text{max}}$, which (seemingly) flattens as $p_{\text{max}}$ increases.}
    \label{prime_SFF}
\end{figure}

Let us consider a hypothetical system whose energy levels are given by  
\[
\omega_0 = \log p_0, \quad \omega_1 = \log p_1, \quad \omega_2 = \log p_2, \ldots
\]  
where \( p_i \) denotes the prime numbers. We will view this as our first quantized system. It turns out that this system in fact already has a weak ramp of slope (much) smaller than 1, if we compute its partition function up to a $p_{max}$. As $p_{max}$ increases, the slope of the ramp appears to flatten. See Figure \ref{prime_SFF}. This is a purely ``experimental'' fact, and we do not have a  deep understanding of it. 

The (bosonic) second quantization of this system is known as the {\em primon gas}. (Even though we have not discussed it above, one can of course also construct a fermionic version of the primon gas where the creation-annihilation operators satisfy the algebra of  fermionic oscillators.)  

Let \( N_0, N_1, N_2, \ldots \) denote the number of bosonic excitations \( \omega_0, \omega_1, \omega_2, \ldots \), respectively. 
The second quantized energy level for a given set \( \{N_i\} \) is  
\begin{align}
    E &= N_0 \log p_0 + N_1 \log p_1 + N_2 \log p_2 + \ldots \nonumber \\
    &= \log p_0^{N_0} + \log p_1^{N_1} + \log p_2^{N_2} + \ldots \nonumber \\
    &= \log (p_0^{N_0} p_1^{N_1} p_2^{N_2} \dots).
\end{align}  
The expression inside the logarithm is simply the well-known \textit{prime factorization} of an integer which means that any number can be uniquely decomposed in this way. For example,  
\begin{equation*}
    84 = 2^2 \cdot 3^1 \cdot 7^1
\end{equation*}  
which implies that the energy \( E = \log 84 \) corresponds to two particles in the state \( \log 2 \), one particle in the state \( \log 3 \), and one particle in the state \( \log 7 \). To sum over all energy levels of the second quantized canonical partition function, we therefore have to sum over al integers $n$ and the corresponding energy level is $\log n$. The primon gas partition function is therefore
\begin{equation}
    Z = \sum_{n} e^{-\beta \log n} = \zeta(\beta), \quad \text{for } \beta > 1.
\end{equation}  
For \( \beta < 1 \), the partition sum diverges, implying that the primon gas cannot be heated to arbitrarily high temperatures. The critical temperature corresponding to \( \beta = 1 \) is what we viewed as a Hagedorn temperature, above which the system no longer behaves like an ordinary gas.

\subsection{Black Hole Microstates from Third Quantization of Primons?}

One of the conclusions of some recent \cite{Pradipta1} and upcoming papers is that low-lying black hole normal modes are approximately logarithmic $\omega_J \sim a + b \log J$. Here $J$ is an angular momentum quantum number, and we are suppressing the radial quantum number $n$ (dependence on it is approximately linear in this regime). Note that the constant piece $a$ can be absorbed into the normalization of the SFF and $b$ simply rescales the temperature. So, as far as studying the SFF is considered, this system is equivalent  to the log spectrum. 

Many interesting features of these modes were noted in \cite{FirstRamp, FuzzRandom, SimplestRamp, Burman1, Pradipta1, Burman2}. The $J$-dependence is weak, so if one ignores it and {\em second} quantizes these modes (i.e., considers the field theory of their parent scalar field) the resulting quantum field theory on the black hole background reproduces both the exterior correlators and thermodynamics of the black hole (including the area law). In terms of the {\em first} quantized normal modes on the other hand, if one is more sophisticated  and retains the $J$-dependence, multiple RMT/chaos aspects of black holes are reproduced -- with various minor qualifications, one finds the linear ramp, level repulsion and (as we saw in the present paper) the $\mathcal{O}(1)$ Thouless time. 

These facts raise the following question: Can we compute the SFF of the second quantized normal mode system while we retain the weak $\sim \log J$ dependence in the spectrum? This would crudely correspond to a third quantization of the primons.  Because the number of states grows very fast, and since SFF requires phase information (unlike the partition function), doing this directly seems to be numerically challenging. But this third quantized partition function and SFF may be interesting objects for future study.

\section{Zoo of $L$-Functions}\label{app:zoo}

The following is a small compilation of some well-known arithmetic $L$-functions~\cite{lmfdb}. This list should be viewed as a brief sampling. We do not provide a detailed discussion of the underlying number theory. The key point for us is that these are explicitly definable functions that can be worked with on Python. We emphasize that the list of arithmetic $L$-functions is huge, see \cite{lmfdb}.

\subsection*{1. The Riemann Zeta Function}
The Riemann zeta function is defined as:  
\[
\zeta(s) = \sum_{n=1}^{\infty} \frac{1}{n^s}, \quad \text{Re}(s) > 1.
\]
It satisfies the Euler product representation:  
\[
\zeta(s) = \prod_p \frac{1}{1 - p^{-s}}.
\]

\subsection*{2. Dirichlet $L$-Functions}
A Dirichlet $L$-function generalizes the Riemann zeta function using a Dirichlet character \( \chi(n) \): 
\begin{equation}\label{DLfn}
    L(s, \chi) = \sum_{n=1}^{\infty} \frac{\chi(n)}{n^s}.
\end{equation}
It has the Euler product:  
\begin{equation*}
L(s, \chi) = \prod_p \left(1 - \frac{\chi(p)}{p^s} \right)^{-1}.
\end{equation*}
In Figure \ref{comparison_DirichletL}, the Dirichlet $L$-functions are represented as \texttt{DirichletL[a,b,s]}, where $\chi$ denotes the $b$-th Dirichlet character modulo $a$.

\subsection*{3. Dedekind Zeta Functions}
For a number field \( K \), the Dedekind zeta function is: 
\begin{equation}\label{DeDefn}
\zeta_K(s) = \sum_{\mathfrak{a}} \frac{1}{N(\mathfrak{a})^s},
\end{equation}
where the sum runs over all nonzero ideals \( \mathfrak{a} \) of the ring of integers \( \mathcal{O}_K \), and \( N(\mathfrak{a}) \) is the norm of \( \mathfrak{a} \). 

In Figure \ref{comparison_Dedekind}, we consider the case $K=\mathbb{Q}(\sqrt{D})$, which represents a quadratic field with radicand $D$.

\subsection*{4. L-Functions of Elliptic Curves}
For an elliptic curve \( E \) (i.e. $y^2=x^3+\mu x+\nu$) over \( \mathbb{Q} \), the $L$-function is defined as:  
\begin{equation}\label{elipeq}
L(E, s) = \sum_{n=1}^{\infty} \frac{a_n}{n^s}.
\end{equation}
For a prime \( p \) where \( E \) has good reduction\footnote{That is, $\Delta\pmod p \equiv -16(4 \mu^3+27 \nu^2)\pmod p\neq 0$.} 
\[
a_p = p + 1 - \#E(\mathbb{F}_p).
\]
Here \( \#E(\mathbb{F}_p) \) is the number of points on the elliptic curve modulo $p$, i.e., the number of solutions to the elliptic curve equation $y=x^3+\mu x+ \nu$ where $x$ and $y$ are taken modulo $p$.
In Figure \ref{comparison_elliptic} these parameters $\mu, \nu$ are represented by $[\mu,\nu]$.



\section{Selberg's Criteria for $L$-Functions}
\label{app:selberg}

The Selberg class $\mathcal{S}$ \cite{cogdell2004lectures}  provides an axiomatic framework for Dirichlet series to be $L$-functions. The key point is that number theory is a natural source of explicitly known $L$-functions\footnote{Examples of such arithmetic $L$-functions were provided in Appendix \ref{app:zoo}.  In this paper, we have not discussed the origins of arithmetic $L$-functions in any detail, rather we have simply chosen to work with some explicit examples.}. Selberg's axioms give a set of {\em analytic} criteria that are conjecturally satisfied by every possible {\em arithmetic} $L$-function. 


A function $L(s)$ belongs to the Selberg class if it satisfies the following axioms:

\begin{enumerate}
    \item \textbf{Analytic Continuation}: The Dirichlet series
    \begin{equation}
        L(s) = \sum_{n=1}^{\infty} \frac{a_n}{n^s}, \quad \text{convergent for } \Re(s)>1,
    \end{equation}
    extends to a meromorphic function on $\mathbb{C}$, with at most a simple pole at $s=1$.

    \item \textbf{Functional Equation}: There exists a completed function
    \begin{equation}
        \Lambda(s) = \omega Q^s \prod_{i=1}^r \Gamma(\lambda_i s + \alpha_i) L(s)
    \end{equation}
    satisfying the reflection symmetry
    \begin{equation}
        \Lambda(s) =  \overline{\Lambda}(1-s),
    \end{equation}
    where $Q>0$, $\lambda_i>0$, $\Re(\alpha_i) \geq 0$, and $|\omega|=1$. Note that in the known arithmetic $L$-functions, these parameters are often more constrained than what is being demanded here\footnote{For example, in known arithmetic $L$-functions, $\lambda_j$ can be taken to be $1/2$ or $1$. Such arithmetic facts have been used in writing the functional equation in the form \eqref{LMFDB-FunEq} that we used in the main body of the paper. Let us also note that $Q$ is related to $N$, $\omega$ is related to $\epsilon$, and $\alpha_i$ are related to the spectral parameters $\mu_j$ and $\nu_k$. In known arithmetic examples, the spectral parameters are real. But our understanding is that this is not required by the Selberg criteria.}.

    \item \textbf{Euler Product}: For $\Re(s)>1$, $L(s)$ has an Euler product expansion:
    \begin{equation}
        L(s) = \prod_p \exp\left(\sum_{k=1}^{\infty} \frac{b_{p^k}}{p^{ks}}\right).
    \end{equation}
    The coefficients $b_{p^k}$ are such that the product converges absolutely. 

    \item \textbf{Ramanujan-Petersson Conjecture}: There exists a constant $C(\epsilon)$ depending only on $\epsilon$ and not $n$, such that the coefficients satisfy  
\begin{equation}
    |a_n| \le C(\epsilon) n^\epsilon, \quad \forall \epsilon > 0.
\end{equation}
\end{enumerate}

The first three of these criteria play fairly direct roles in our discussions. 

\section{Truncated $L$-functions and Plateau}\label{truncated_L}

\begin{figure}[ht]
    \centering
    \begin{subfigure}{0.48\textwidth}
        \centering
        \includegraphics[width=\linewidth]{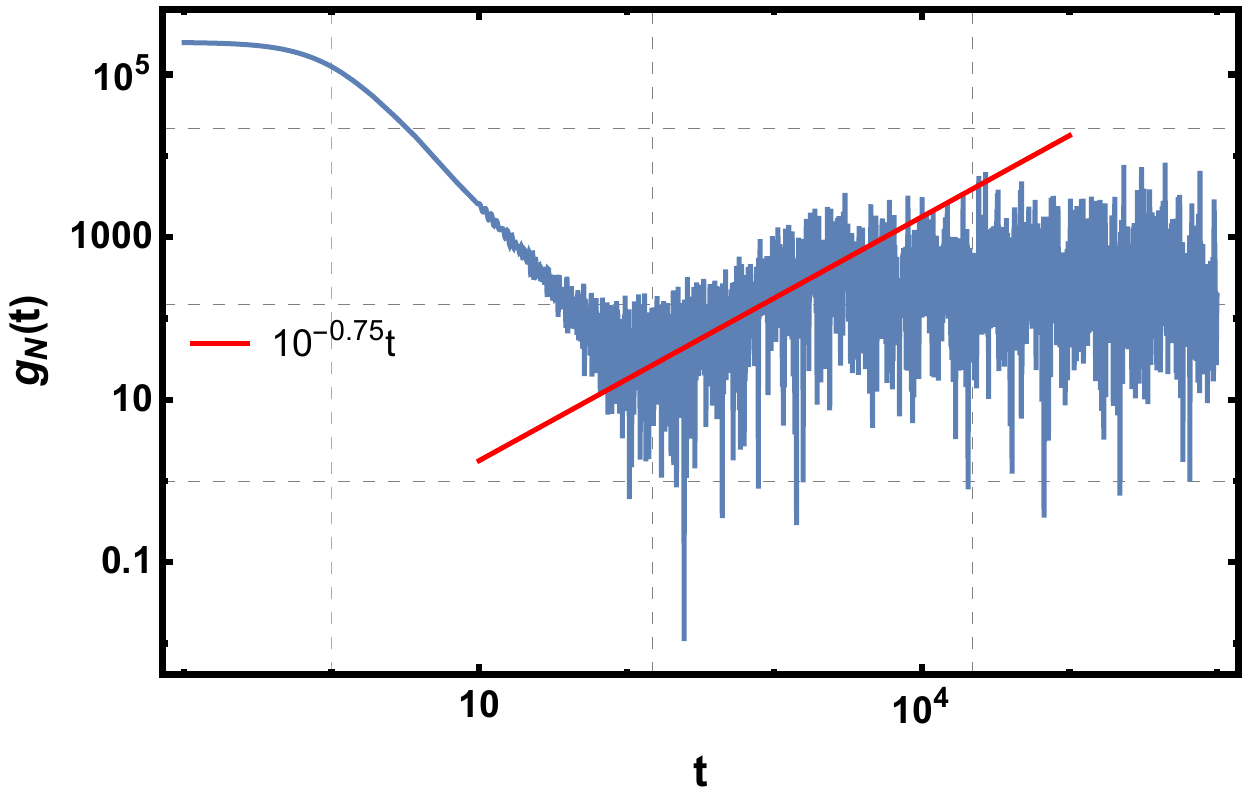}
        \caption{$N=1000$}
    \end{subfigure}
    \hfill
    \begin{subfigure}{0.48\textwidth}
        \centering
        \includegraphics[width=\linewidth]{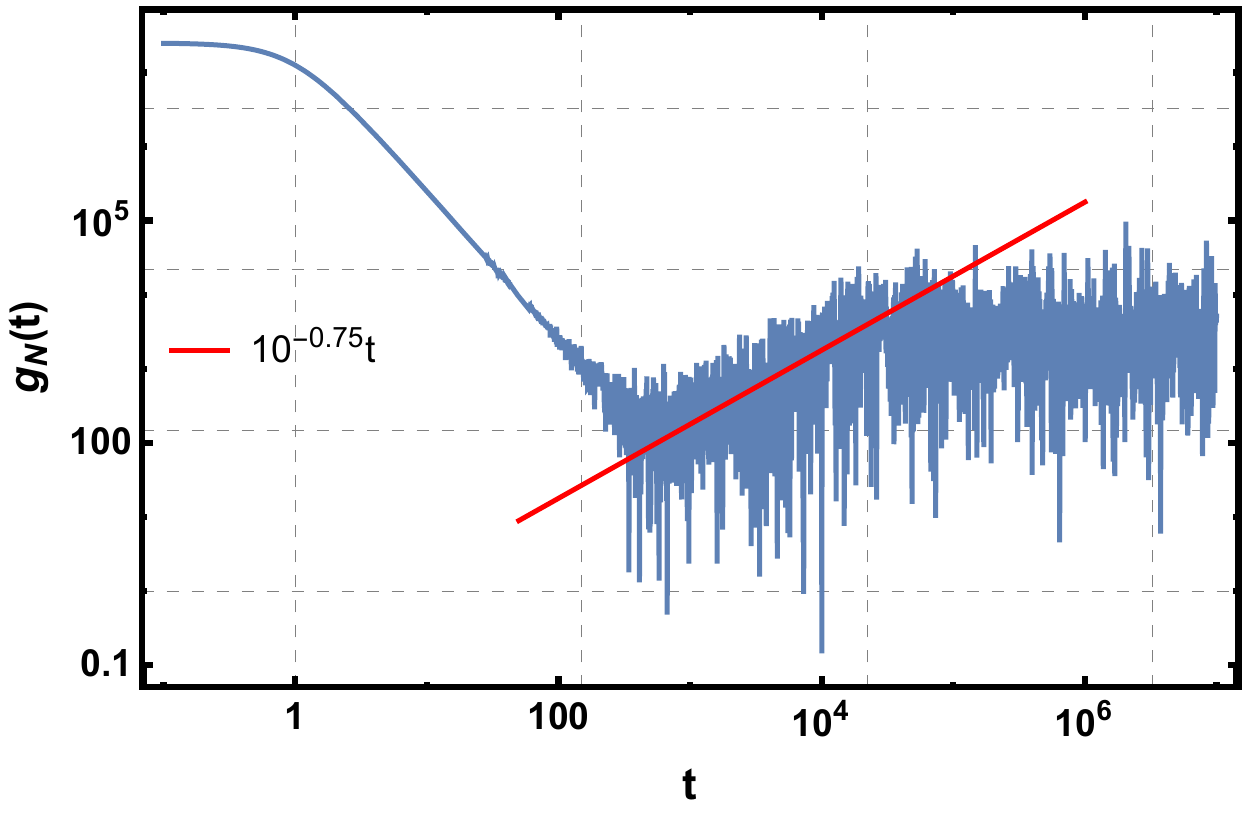}
        \caption{$N=10000$}
    \end{subfigure}
    \caption{The SFF $g_N(t)$ constructed from the truncated Dirichlet $L$-series, \texttt{DirichletL[2,1,0 + it]}, with truncation at $n=N$ and evaluated at $\beta=0$. The plots include a red line of unit slope to highlight the linear ramp in the SFF. Truncation induces a plateau at late times.}
    \label{truncatedSFF}
\end{figure}

\begin{figure}[ht]
    \centering
    \begin{subfigure}{0.48\textwidth}
        \centering
        \includegraphics[width=\linewidth]{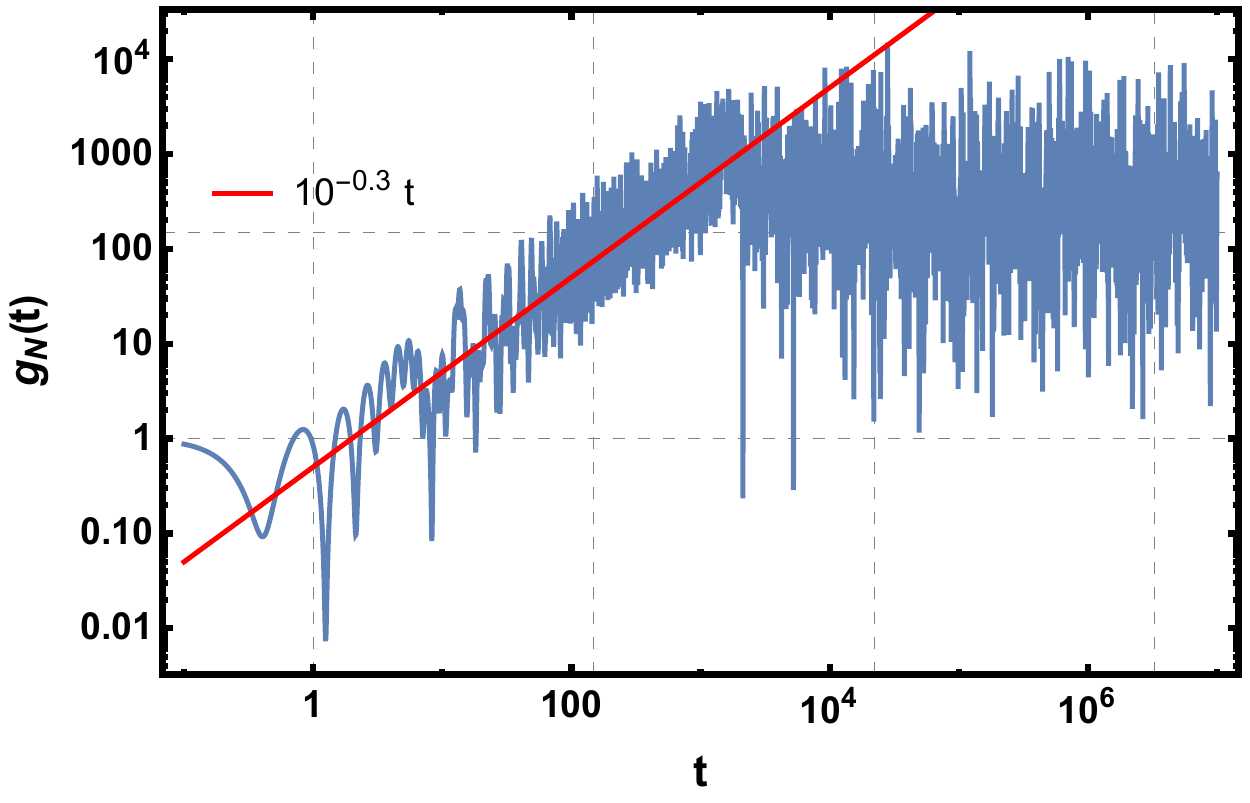}
        \caption{Unnormalized}
    \end{subfigure}
    \hfill
    \begin{subfigure}{0.48\textwidth}
        \centering
        \includegraphics[width=\linewidth]{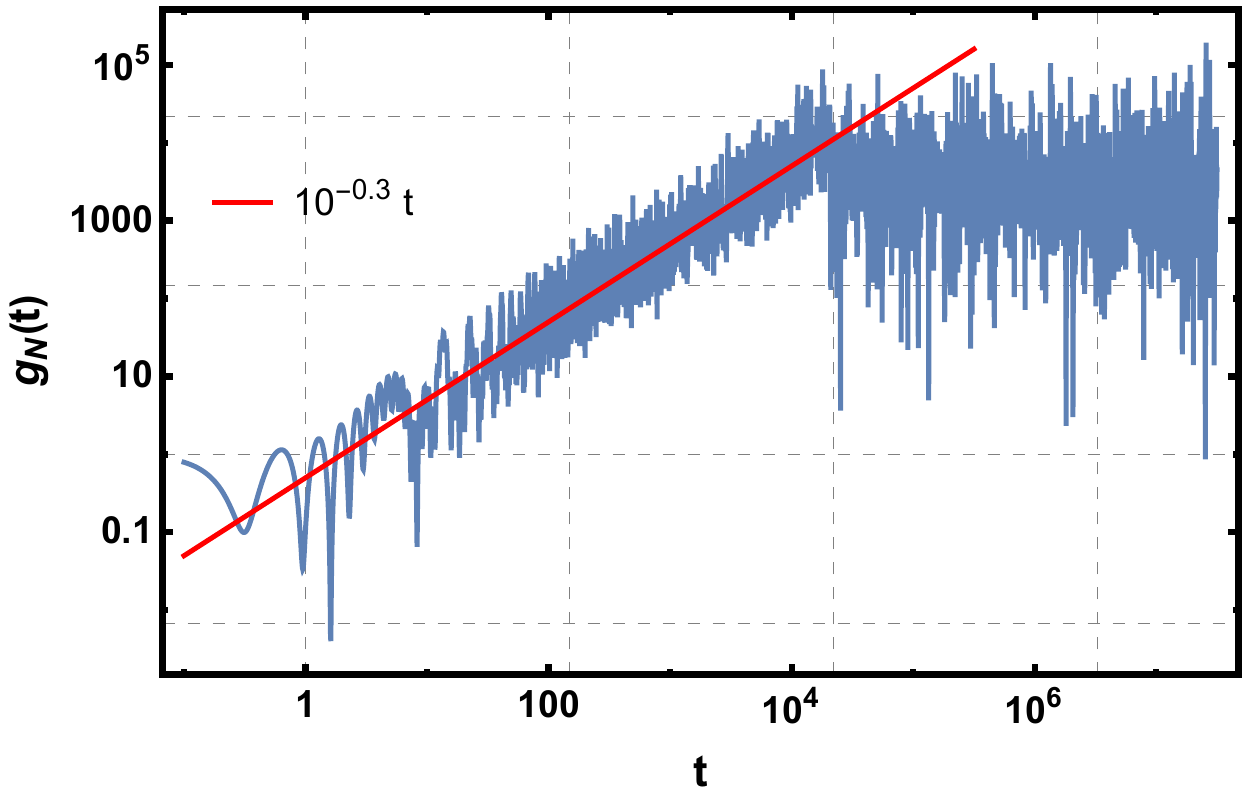}
        \caption{Normalized ($N=10000$)}
    \end{subfigure}
    \caption{The SFF $g_N(t)$ obtained from a truncated Dirichlet $L$-function \texttt{DirichletL[3,2,$it$]} with truncation at $N$ and evaluated at $\beta=0$. Here also the red line with unit slope illustrates the linear growth characteristic of the ramp.}
    \label{SFF_L32x}
\end{figure}

In the main text, we observed that the SFF constructed from various $L$-functions exhibits an eternal ramp above the Hagedorn temperature with a definite slope, but no plateau, like the case of the zeta function. To recover a plateau in this regime of $\beta$, we truncate the infinite sum at some $n=N$
\begin{equation}\label{Lfn_truncated}
    L_N(s, \chi) = \sum_{n=1}^{N} \frac{\chi(n)}{n^s}.
\end{equation}
The SFF takes the form:
\begin{equation}
    g_N(t) = |L_N(it, \chi_j)|^2 .
\end{equation}
Figure \ref{truncatedSFF} and \ref{SFF_L32x} show the SFF at $\beta=0$, displaying the expected dip-ramp-plateau structure. A red line with unit slope is also included to highlight that the ramp's slope is $\sim 1$.

\section{Approximations of Gamma Factors}
Stirling's Approximation:
\begin{align*}
&\Gamma(z) \sim \sqrt{\frac{2\pi}{z}}\left(\frac{z}{e}\right)^z \quad \text{for } |z| \to \infty \text{ with } |\arg z| < \pi \\[2em]
\end{align*}
Gamma Ratio Approximation \cite{DLMF}:
\begin{align}\label{GammaRatio}
&\frac{\Gamma(z+a)}{\Gamma(z)} \sim z^a\quad \text{for } |z| \to \infty \text{ with } |\arg z| < \pi
\end{align}

\bibliography{bibliography1.bib}

\providecommand{\href}[2]{#2}\begingroup\raggedright\begin{thebibliography}{10}

\bibitem{Vishveshwara:1970zz}
C.V.~Vishveshwara, \emph{{Scattering of Gravitational Radiation by a
  Schwarzschild Black-hole}},
  \href{https://doi.org/10.1038/227936a0}{\emph{Nature} {\bfseries 227} (1970)
  936}.

\bibitem{tHooft}
G.~'t~Hooft, \emph{{On the Quantum Structure of a Black Hole}},
  \href{https://doi.org/10.1016/0550-3213(85)90418-3}{\emph{Nucl. Phys. B}
  {\bfseries 256} (1985) 727}.

\bibitem{FirstRamp}
S.~Das, C.~Krishnan, A.P.~Kumar and A.~Kundu, \emph{{Synthetic fuzzballs: a
  linear ramp from black hole normal modes}},
  \href{https://doi.org/10.1007/JHEP01(2023)153}{\emph{JHEP} {\bfseries 01}
  (2023) 153} [\href{https://arxiv.org/abs/2208.14744}{{\ttfamily
  2208.14744}}].

\bibitem{FuzzRandom}
S.~Das, S.K.~Garg, C.~Krishnan and A.~Kundu, \emph{{Fuzzballs and random
  matrices}}, \href{https://doi.org/10.1007/JHEP10(2023)031}{\emph{JHEP}
  {\bfseries 10} (2023) 031}
  [\href{https://arxiv.org/abs/2301.11780}{{\ttfamily 2301.11780}}].

\bibitem{SimplestRamp}
S.~Das, S.K.~Garg, C.~Krishnan and A.~Kundu, \emph{{What is the Simplest Linear
  Ramp?}}, \href{https://doi.org/10.1007/JHEP01(2024)172}{\emph{JHEP}
  {\bfseries 01} (2024) 172}
  [\href{https://arxiv.org/abs/2308.11704}{{\ttfamily 2308.11704}}].

\bibitem{SumanArnab}
S.~Das and A.~Kundu, \emph{{Brickwall in rotating BTZ: a dip-ramp-plateau
  story}}, \href{https://doi.org/10.1007/JHEP02(2024)049}{\emph{JHEP}
  {\bfseries 02} (2024) 049}
  [\href{https://arxiv.org/abs/2310.06438}{{\ttfamily 2310.06438}}].

\bibitem{Pradipta1}
C.~Krishnan and P.S.~Pathak, \emph{{Normal modes of the stretched horizon: a
  bulk mechanism for black hole microstate level spacing}},
  \href{https://doi.org/10.1007/JHEP03(2024)162}{\emph{JHEP} {\bfseries 03}
  (2024) 162} [\href{https://arxiv.org/abs/2312.14109}{{\ttfamily
  2312.14109}}].

\bibitem{Burman1}
V.~Burman, S.~Das and C.~Krishnan, \emph{{A smooth horizon without a smooth
  horizon}}, \href{https://doi.org/10.1007/JHEP03(2024)014}{\emph{JHEP}
  {\bfseries 2024} (2024) 014}
  [\href{https://arxiv.org/abs/2312.14108}{{\ttfamily 2312.14108}}].

\bibitem{Souvik}
S.~Banerjee, S.~Das, M.~Dorband and A.~Kundu, \emph{{Brickwall, normal modes,
  and emerging thermality}},
  \href{https://doi.org/10.1103/PhysRevD.109.126020}{\emph{Phys. Rev. D}
  {\bfseries 109} (2024) 126020}
  [\href{https://arxiv.org/abs/2401.01417}{{\ttfamily 2401.01417}}].

\bibitem{Ranjini}
C.~Krishnan and R.~Mondol, \emph{{Young Black Holes Have Smooth Horizons: A
  Swampland Argument}},  \href{https://arxiv.org/abs/2407.11952}{{\ttfamily
  2407.11952}}.

\bibitem{Suchetan}
S.~Das, \emph{{Stretched horizon from conformal field theory}},
  \href{https://doi.org/10.1007/JHEP11(2024)033}{\emph{JHEP} {\bfseries 11}
  (2024) 033} [\href{https://arxiv.org/abs/2406.10879}{{\ttfamily
  2406.10879}}].

\bibitem{Burman2}
V.~Burman and C.~Krishnan, \emph{{A Bottom-Up Approach to Black Hole
  Microstates}},  \href{https://arxiv.org/abs/2409.05850}{{\ttfamily
  2409.05850}}.

\bibitem{SumanBaishali}
S.~Das, S.~Porey and B.~Roy, \emph{{Brick Wall in AdS-Schwarzschild Black Hole:
  Normal Modes and Emerging Thermality}},
  \href{https://arxiv.org/abs/2409.05519}{{\ttfamily 2409.05519}}.

\bibitem{Souvik2}
S.~Banerjee, S.~Das, A.~Kundu and M.~Sittinger, \emph{{Blackish Holes}},
  \href{https://arxiv.org/abs/2411.09500}{{\ttfamily 2411.09500}}.

\bibitem{Pradipta2}
C.~Krishnan and P.S.~Pathak, \emph{{Holomorphic Factorization at the Quantum
  Horizon}},  \href{https://arxiv.org/abs/2410.00732}{{\ttfamily 2410.00732}}.

\bibitem{Cotler}
J.S.~Cotler, G.~Gur-Ari, M.~Hanada, J.~Polchinski, P.~Saad, S.H.~Shenker
  et~al., \emph{{Black Holes and Random Matrices}},
  \href{https://doi.org/10.1007/JHEP05(2017)118}{\emph{JHEP} {\bfseries 05}
  (2017) 118} [\href{https://arxiv.org/abs/1611.04650}{{\ttfamily
  1611.04650}}].

\bibitem{ShenkerBPS}
Y.~Chen, H.W.~Lin and S.H.~Shenker, \emph{{BPS chaos}},
  \href{https://doi.org/10.21468/SciPostPhys.18.2.072}{\emph{SciPost Phys.}
  {\bfseries 18} (2025) 072}
  [\href{https://arxiv.org/abs/2407.19387}{{\ttfamily 2407.19387}}].

\bibitem{Mehta1960ONTS}
M.L.~Mehta, \emph{On the statistical properties of the level-spacings in
  nuclear spectra}, {\emph{Nuclear Physics} {\bfseries 18} (1960) 395}.

\bibitem{dyson1962statistical}
F.J.~Dyson, \emph{Statistical theory of the energy levels of complex systems.
  iii}, {\emph{Journal of Mathematical Physics} {\bfseries 3} (1962) 166}.

\bibitem{JuliaNumbers}
B.~Julia, \emph{{STATISTICAL THEORY OF NUMBERS}},  in \emph{{Les Houches School
  of Theoretical Physics: Number Theory and Physics}}, 9, 1989.

\bibitem{BakasGases}
I.~Bakas and M.J.~Bowick, \emph{{Curiosities of arithmetic gases}},
  \href{https://doi.org/10.1063/1.529511}{\emph{J. Math. Phys.} {\bfseries 32}
  (1991) 1881}.

\bibitem{witten1998antidesitterspacethermal}
E.~Witten, \emph{Anti-de sitter space, thermal phase transition, and
  confinement in gauge theories},  1998.

\bibitem{Aharony:2003sx}
O.~Aharony, J.~Marsano, S.~Minwalla, K.~Papadodimas and M.~Van~Raamsdonk,
  \emph{{The Hagedorn - deconfinement phase transition in weakly coupled large
  N gauge theories}},
  \href{https://doi.org/10.4310/ATMP.2004.v8.n4.a1}{\emph{Adv. Theor. Math.
  Phys.} {\bfseries 8} (2004) 603}
  [\href{https://arxiv.org/abs/hep-th/0310285}{{\ttfamily hep-th/0310285}}].

\bibitem{Sundborg:1999ue}
B.~Sundborg, \emph{{The Hagedorn transition, deconfinement and N=4 SYM
  theory}}, \href{https://doi.org/10.1016/S0550-3213(00)00044-4}{\emph{Nucl.
  Phys. B} {\bfseries 573} (2000) 349}
  [\href{https://arxiv.org/abs/hep-th/9908001}{{\ttfamily hep-th/9908001}}].

\bibitem{Titchmarsh}
E.C.~Titchmarsh, \emph{The Theory of the Riemann Zeta-Function}, Oxford
  University Press, second edition, revised by d. r. heath-brown~ed. (1986).

\bibitem{HanadaShenker}
H.~Gharibyan, M.~Hanada, S.H.~Shenker and M.~Tezuka, \emph{{Onset of Random
  Matrix Behavior in Scrambling Systems}},
  \href{https://doi.org/10.1007/JHEP07(2018)124}{\emph{JHEP} {\bfseries 07}
  (2018) 124} [\href{https://arxiv.org/abs/1803.08050}{{\ttfamily
  1803.08050}}].

\bibitem{Dario}
T.~Nosaka, D.~Rosa and J.~Yoon, \emph{{The Thouless time for mass-deformed
  SYK}}, \href{https://doi.org/10.1007/JHEP09(2018)041}{\emph{JHEP} {\bfseries
  09} (2018) 041} [\href{https://arxiv.org/abs/1804.09934}{{\ttfamily
  1804.09934}}].

\bibitem{murty2024lfunction}
M.R.~Murty, ``What is an l-function?.''
  \url{https://www.youtube.com/watch?v=922Jsbg487A}, September, 2024.

\bibitem{peakmath_channel}
A.~Holmström and the PeakMath~Team, ``Peakmath landscape.''
  \url{https://www.youtube.com/@PeakMathLandscape}, 2023.

\bibitem{IwaniecKowalski}
H.~Iwaniec and E.~Kowalski, \emph{Analytic Number Theory}, vol.~53 of \emph{AMS
  Colloquium Publications}, American Mathematical Society (2004).

\bibitem{MontgomeryVaughan}
H.L.~Montgomery and R.C.~Vaughan, \emph{Multiplicative Number Theory I:
  Classical Theory}, Cambridge University Press (2007).

\bibitem{lmfdb}
{The LMFDB Collaboration}, ``The {L}-functions and modular forms database.''
  \url{https://www.lmfdb.org}, 2025.

\bibitem{Gubser_2017}
S.S.~Gubser, J.~Knaute, S.~Parikh, A.~Samberg and P.~Witaszczyk, \emph{p-adic
  ads/cft},
  \href{https://doi.org/10.1007/s00220-016-2813-6}{\emph{Communications in
  Mathematical Physics} {\bfseries 352} (2017) 1019–1059}.

\bibitem{Maldacena_2016}
J.~Maldacena, S.H.~Shenker and D.~Stanford, \emph{A bound on chaos},
  \href{https://doi.org/10.1007/jhep08(2016)106}{\emph{Journal of High Energy
  Physics} {\bfseries 2016} (2016) }.

\bibitem{cogdell2004lectures}
J.~Cogdell, H.~Kim and M.~Murty, \emph{Lectures on Automorphic L-functions},
  Fields Institute monographs, American Mathematical Society (2004).

\bibitem{DLMF}
{NIST Digital Library of Mathematical Functions}, ``Gamma function properties:
  Asymptotic expansions.'' \url{https://dlmf.nist.gov/5.11}, 2025.

\end{thebibliography}\endgroup
\bibliographystyle{JHEP.bst}

\end{document}